\documentclass[prd,preprint,showpacs,preprintnumbers,nofootinbib,eqsecnum,superscriptaddress]{revtex4}
\pdfoutput=1

 \usepackage[dvips,final]{graphicx}
  \usepackage{amssymb}
   \usepackage{amsmath}
   \usepackage{hyperref}
    \usepackage{amsfonts}
     \usepackage{epsfig}
      \usepackage{bm}

\usepackage{mathpazo}

\usepackage[section]{placeins}

\usepackage{multirow}
\usepackage{ctable}
\usepackage{booktabs}
\usepackage{array}
\usepackage{tabularx}
\usepackage{xcolor}
\usepackage{pstricks}

\newcommand{\nn}{\nonumber}

\newcommand{\beq}{\begin{equation}}
\newcommand{\eeq}{\end{equation}}
\newcommand{\bea}{\begin{eqnarray}}
\newcommand{\eea}{\end{eqnarray}}
\newcommand{\beqa}{\begin{eqnarray}}
\newcommand{\eeqa}{\end{eqnarray}}

\definecolor{red}{rgb}{1,0,0}

\def\be{\begin{equation}}
\def\ee{\end{equation}}
\numberwithin{equation}{section}

\begin{document}

\title{Search for optimal conditions for exploring double-parton scattering\\
in four-jet production: $k_T$-factorization approach}

\author{Krzysztof Kutak}
\email{krzysztof.kutak@ifj.edu.pl} \affiliation{Institute of Nuclear
Physics, Polish Academy of Sciences, Radzikowskiego 152, PL-31-342 Krak{\'o}w, Poland}

\author{Rafa{\l} Maciu{\l}a}
\email{rafal.maciula@ifj.edu.pl} \affiliation{Institute of Nuclear
Physics, Polish Academy of Sciences, Radzikowskiego 152, PL-31-342 Krak{\'o}w, Poland}

\author{Mirko Serino}
\email{mirko.serino@ifj.edu.pl} \affiliation{Institute of Nuclear
Physics, Polish Academy of Sciences, Radzikowskiego 152, PL-31-342 Krak{\'o}w, Poland}

\author{Antoni Szczurek}
\email{antoni.szczurek@ifj.edu.pl} \affiliation{Institute of Nuclear
Physics, Polish Academy of Sciences, Radzikowskiego 152, PL-31-342 Krak{\'o}w, Poland}
\affiliation{University of Rzesz\'ow, PL-35-959 Rzesz\'ow, Poland}

\author{Andreas van Hameren}
\email{hameren@ifj.edu.pl} \affiliation{Institute of Nuclear
Physics, Polish Academy of Sciences, Radzikowskiego 152, PL-31-342 Krak{\'o}w, Poland}

\date{\today}

\begin{abstract}
In the present paper we discuss how to maximize the double-parton
scattering (DPS) contribution in four-jet production by selecting kinematical cuts. Here both
single-parton and double-parton scattering effects are calculated
in the $k_T$-factorization approach, following our recent developments
of relevant methods and tools. Several differential distributions
are shown and discussed in the context of future searches for DPS
effects, such as rapidity of jets, rapidity distance, and azimuthal correlations between jets.
The dependences of the relative DPS amount is studied as function of those observables. 
The regions with an enhanced DPS contribution are identified. 
Future experimental explorations could extract more precise values of $\sigma_{eff}$ and its potential dependence on kinematical variables.
\end{abstract}

\thispagestyle{empty}
\vspace{-28em}
\begin{flushright}
 IFJPAN-IV-2016-14 \\
\end{flushright}
\vspace{2em}
\maketitle

\section{Introduction}

The relative amount of hard double-parton scattering (DPS) grows with energy. 
This is because the density of partons (sea quarks and antiquarks and gluons) 
grows with decreasing values of the longitudinal momentum fractions $x_1$ and $x_2$ 
of the first and second hadron momenta, respectively.
The larger the energy, the smaller the values of the longitudinal momentum 
fractions necessary for hard scattering to take place are. 

This is particularly true for processes induced by gluon-gluon fusion,
like charm production for instance \cite{Luszczak:2011zp,vanHameren:2014ava,Maciula:2016wci}.
So far, most practical calculations of DPS contributions were performed 
within the so-called factorized ansatz often called pocket-formula. 
In this approach, the (differential) cross section for DPS is a product 
of the corresponding (differential) cross sections for single-parton scatterings (SPS). 
This is an approximation which is not well under control yet. 
A better formalism exists in principle, but predictions are not easy, 
as they require unknown input(s) related to the correlation of partons
in configuration space, spin, etc \cite{Diehl:2011yj}.
The latter are explored to a far lesser extent than single-parton distributions. 
In this situation we may try to explore the problem by first collecting
a sufficient amount of empirical facts to draw practical conclusions. 
As proposed by two of us some time ago, double $c \bar c$ production
is a good place to explore DPS \cite{Luszczak:2011zp}.
A new analysis shows that even there the situation may be not that simple \cite{Maciula:2016wci}. 
Four-jet production seems a natural case to look for hard DPS effects \cite{Paver:1983hi,Domdey:2009bg,Berger:2009cm,Blok:2010ge,Maciula:2015vza}.

A year ago two of us analyzed how to find optimal
conditions for the observation and exploration of DPS 
effects in four-jet production \cite{Maciula:2015vza}. In those analyses only 
the leading-order (LO) approach was applied both to SPS and DPS. 
It is expected that higher order effects are provided, already 
at tree level, by the $k_T$-factorization approach. At high energy
the small-x values region opens up which is a further motivation to apply this approach.

Very recently, we have performed for the first time a calculation of 
four-jet production for both single-parton and double-parton mechanism 
within $k_T$-factorization \cite{Kutak:2016mik}. 
It was shown that the effective inclusion of higher-order effects leads 
to a substantial damping of the double-scattering contribution 
with respect to the SPS one, especially for symmetric (identical) cuts on the transverse momenta of all jets. 
In a leading-order approach to $2\to2$ processes, the transverse momenta of the final state jets must have the same size.
Either of them passing the cut automatically implies that the second one is accepted too.
The situation is subtler in the next-to-leading order (NLO) collinear approach or in the tree-level $k_T$-factorization approach,
which is the reason why, as will be recalled also in this paper, asymmetric cuts should do a better job in searches for DPS.
For the purpose of the present analysis, we will take into account higher-order virtual 
effects via K-factors deduced from NLO calculation in collinear factorization.

As discussed in Ref.~\cite{Maciula:2015vza}, jets with a large rapidity separation 
seem more promising than others in exploring DPS effects in four-jet production.
In the following we shall concentrate on the study of this and other more optimal observables
to pin down DPS in the  $k_T$-factorization framework.
Obviously, low cuts on the transverse momenta of jets favour DPS.

\section{A sketch of the theoretical formalism}

We will briefly recall the theoretical formalism we use to obtain our predictions.
This has already been discussed extensively in Ref.~\cite{Kutak:2016mik}, to which we refer for further details,
including references, on both the Transverse Momentum Dependent parton distribution functions (TMDs) 
and the scattering amplitudes with off-shell initial state partons.

The high-energy-factorization (HEF) \cite{Catani:1990eg} formula for the calculation of the inclusive partonic four-jet cross section at the Born level reads 
\bea
\sigma^B_{4-jets} 
&=& 
\sum_{i,j} \int \frac{dx_1}{x_1}\,\frac{dx_2}{x_2}\, d^2 k_{T1} d^2 k_{T2}\,  \mathcal{F}_i(x_1,k_{T1},\mu_F)\, \mathcal{F}_j(x_2,k_{T2},\mu_F) \nn \\
&&
\hspace{-25mm}
\times \frac{1}{2 \hat{s}} \prod_{l=i}^4 \frac{d^3 k_l}{(2\pi)^3 2 E_l} \Theta_{4-jet} \, (2\pi)^4\, \delta\left( x_1P_1 + x_2P_2 + \vec{k}_{T\,1}+ \vec{k}_{T\,2} - \sum_{l=1}^4 k_i \right)\, 
\overline{ \left| \mathcal{M}(i^*,j^* \rightarrow 4\, \text{part.})
\right|^2 } \, . \nn \\
\label{kt_cross}
\eea
Here $\mathcal{F}_i(x_k,k_{Tk},\mu_F)$ is the TMD for a given parton ($k$ numbers the parton type), $x_k$ are the longitudinal
momentum fractions, $\mu_F$ is a factorization scale, $\vec{k}_{Tk}$ are the parton's transverse momenta, perpendicular to the collision axis. In the calculations we use the DLC2016v2 TMD set 
\cite{Kutak:2016mik}\footnote{Available by request from krzysztof.kutak@ifj.edu.pl. The difference to the DLC2016 set is due to promoting running coupling to NLO accuracy. This change affected slightly the observables we study as compared to \cite{Kutak:2016mik}.}.
$\mathcal{M}(i^*,j^* \rightarrow 4\, \text{part.})$ is the gauge invariant matrix element for $2\rightarrow 4$ particle scattering with two initial off-shell legs.
They are evaluated numerically with AVHLIB~\cite{Bury:2015dla}, which also provides the other necessary Monte Carlo tools for the calculation.
In the calculation, the scales are set to
$\mu_F=\mu_R= \frac{\hat{H}_T}{2} = \frac{1}{2} \sum_{l=1}^4 k_T^l$\footnote{As customary in the literature, 
we use the $\hat{H}_T$ notation to refer to the energies of the final state partons, 
not jets, despite this is obviously the same in the LO analysis.}, and we use the $n_F = 5$ flavour scheme.

The so-called pocket-formula for DPS cross sections (for a four-parton final state) is given by
\beq
\frac{d \sigma^{B}_{4-jet,DPS}}{d \xi_1 d \xi_2} = 
\frac{m}{\sigma_{eff}} \sum_{i_1,j_1,k_1,l_1;i_2,j_2,k_2,l_2} 
\frac{d \sigma^B(i_1 j_1 \rightarrow k_1 l_1)}{d \xi_1}\, \frac{d \sigma^B(i_2 j_2 \rightarrow k_2 l_2)}{d \xi_2} \, ,
\label{pocket}
\eeq
where the $\sigma(a b \rightarrow c d)$ cross sections are obtained by
restricting (\ref{kt_cross}) to a single channel and the symmetry factor $m$ is $1/2$ if the two hard
scatterings are identical, to prevent double counting them.
Finally, $\xi_1$ and $\xi_2$ stand for generic kinematical variables for the first and second scattering, respectively.
It goes without saying that such a formula is a phenomenology-motivated approximation.
The effective cross section $\sigma_{eff}$ can be loosely interpreted as a measure of the transverse correlation of the two partons inside 
the hadrons, whereas the possible longitudinal correlations are usually neglected.
As for our previous paper \cite{Kutak:2016mik}, we use  the value $\sigma_{eff}$ = 15 mb, 
although this value may be questioned \cite{Maciula:2016wci} when all SPS mechanisms of double charm production are included.
For recent developments in the formal theory of DPS in the collinear factorization framework, 
we refer the interested reader to \cite{Diehl:2015bca}.

\section{Detailed studies}

\subsection{Comparison to the CMS data}

\begin{figure}[!h]
\begin{minipage}{0.47\textwidth}
 \centerline{\includegraphics[width=1.0\textwidth]{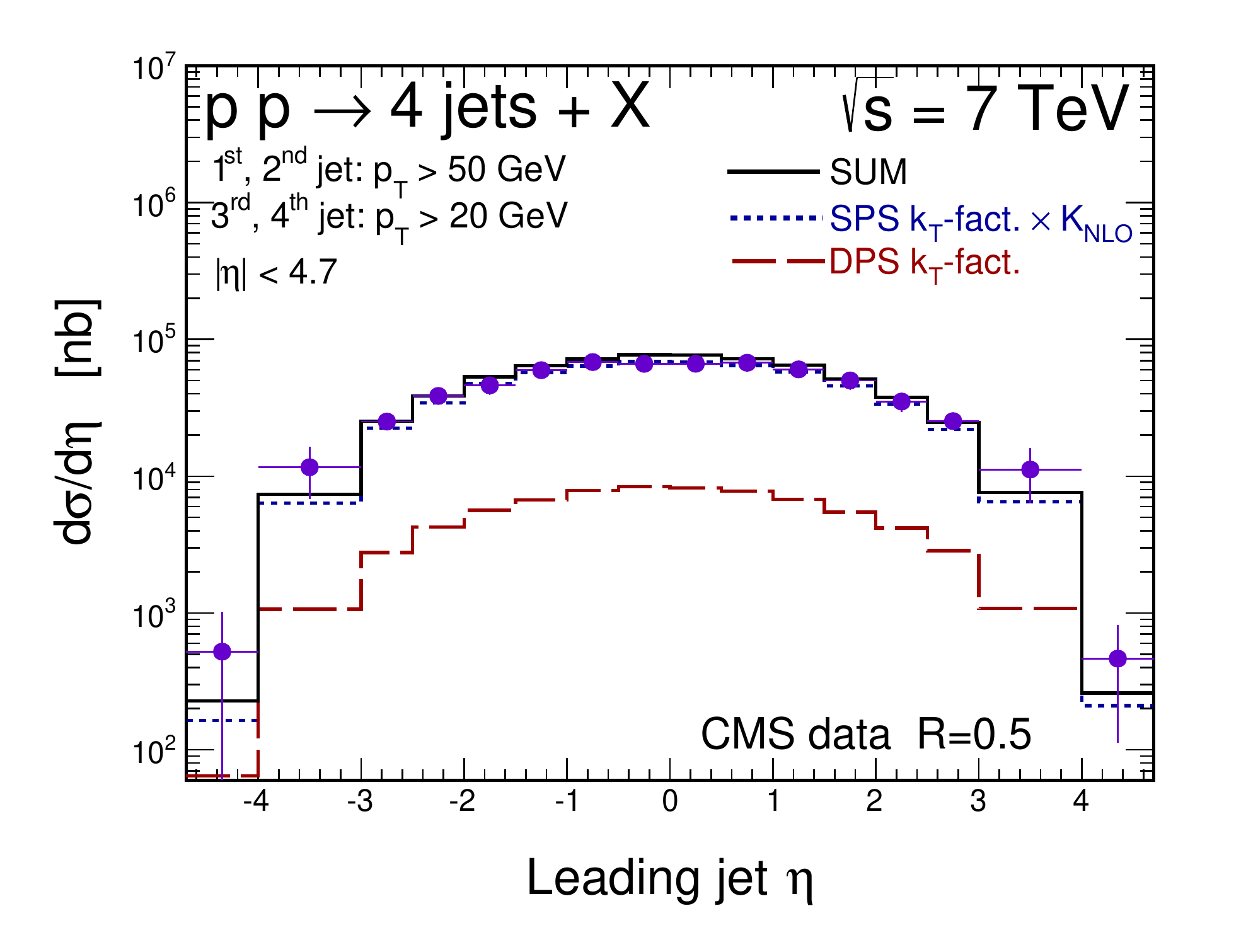}}
\end{minipage}
\hspace{0.5cm}
\begin{minipage}{0.47\textwidth}
 \centerline{\includegraphics[width=1.0\textwidth]{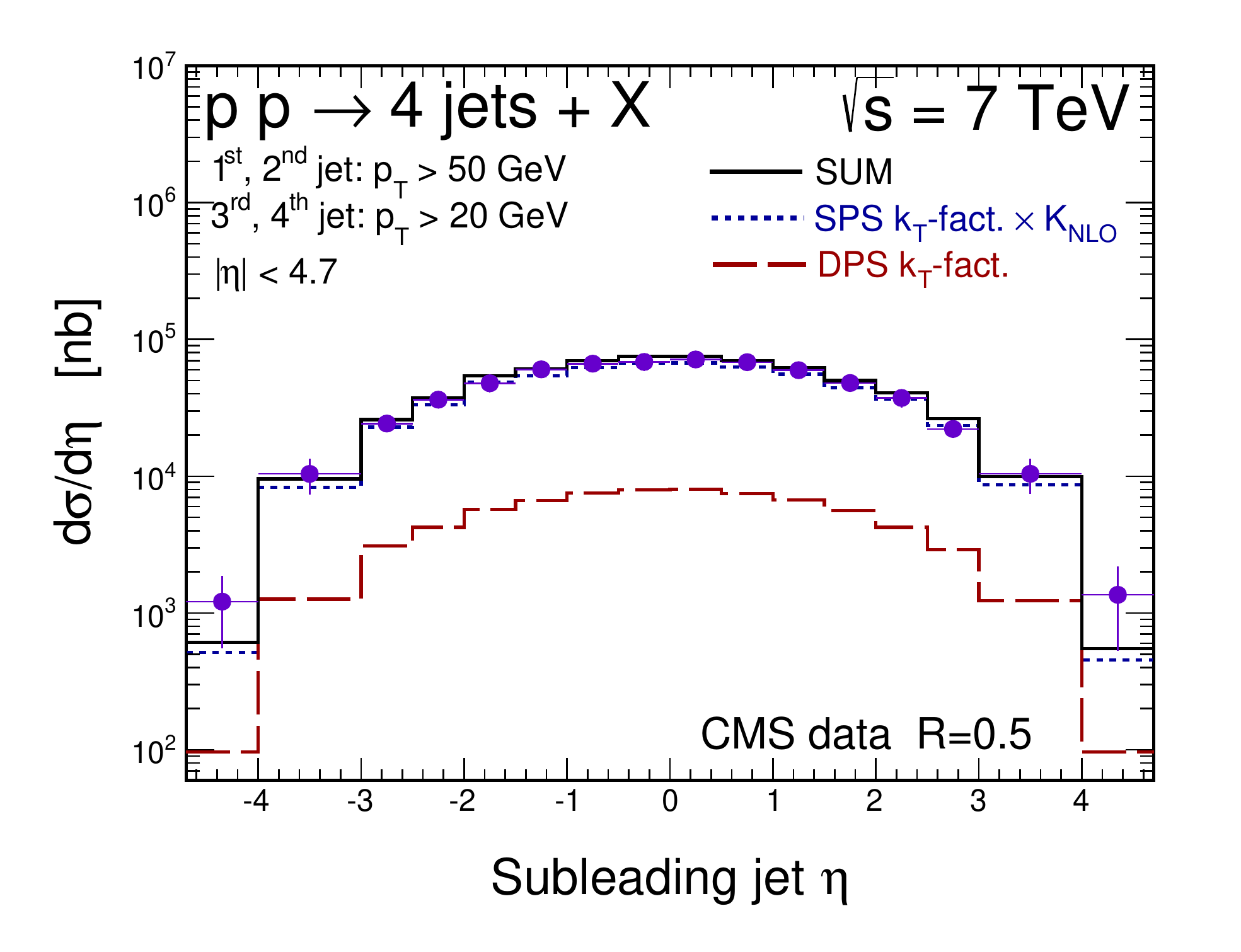}}
\end{minipage}\\
\begin{minipage}{0.47\textwidth}
 \centerline{\includegraphics[width=1.0\textwidth]{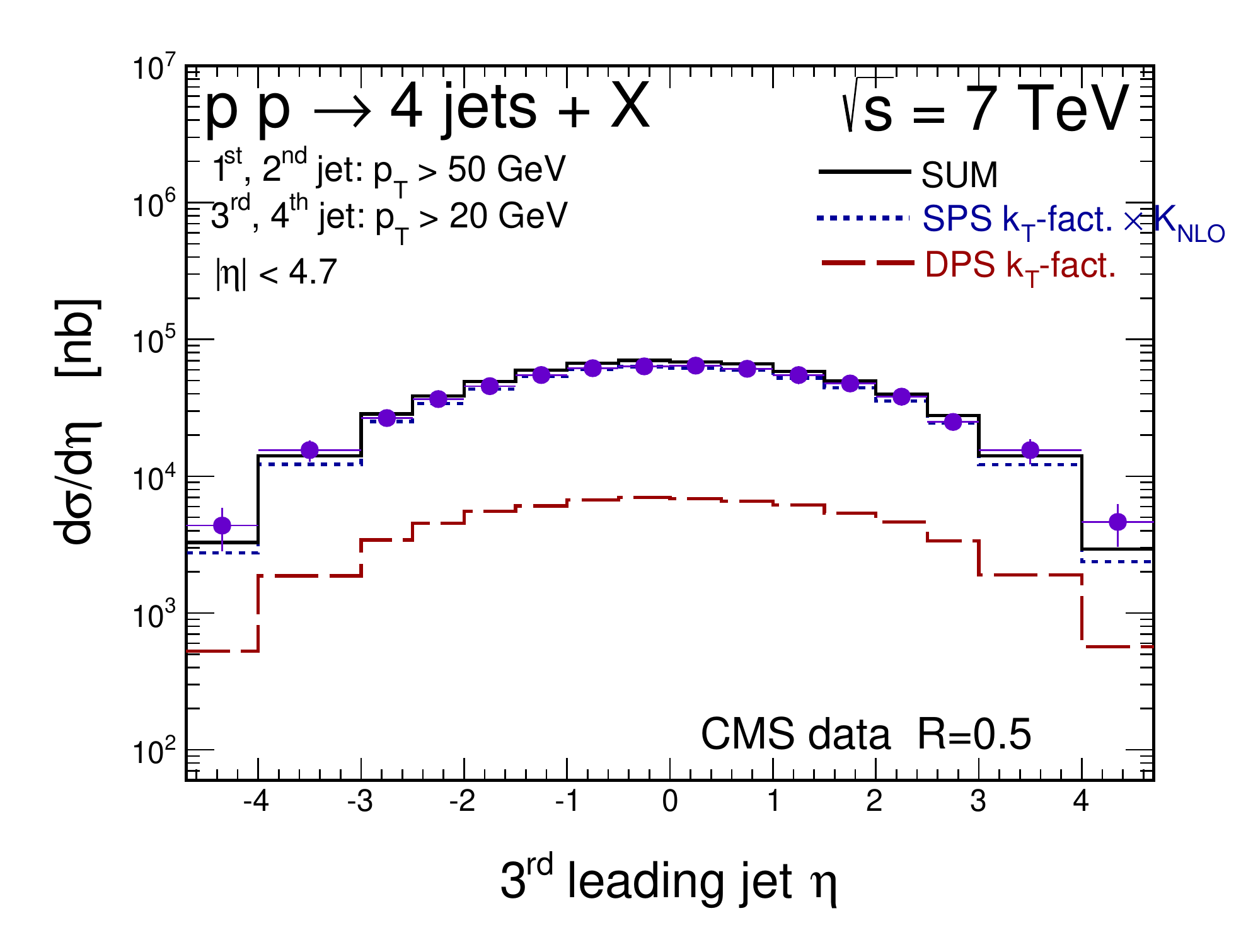}}
\end{minipage}
\hspace{0.5cm}
\begin{minipage}{0.47\textwidth}
 \centerline{\includegraphics[width=1.0\textwidth]{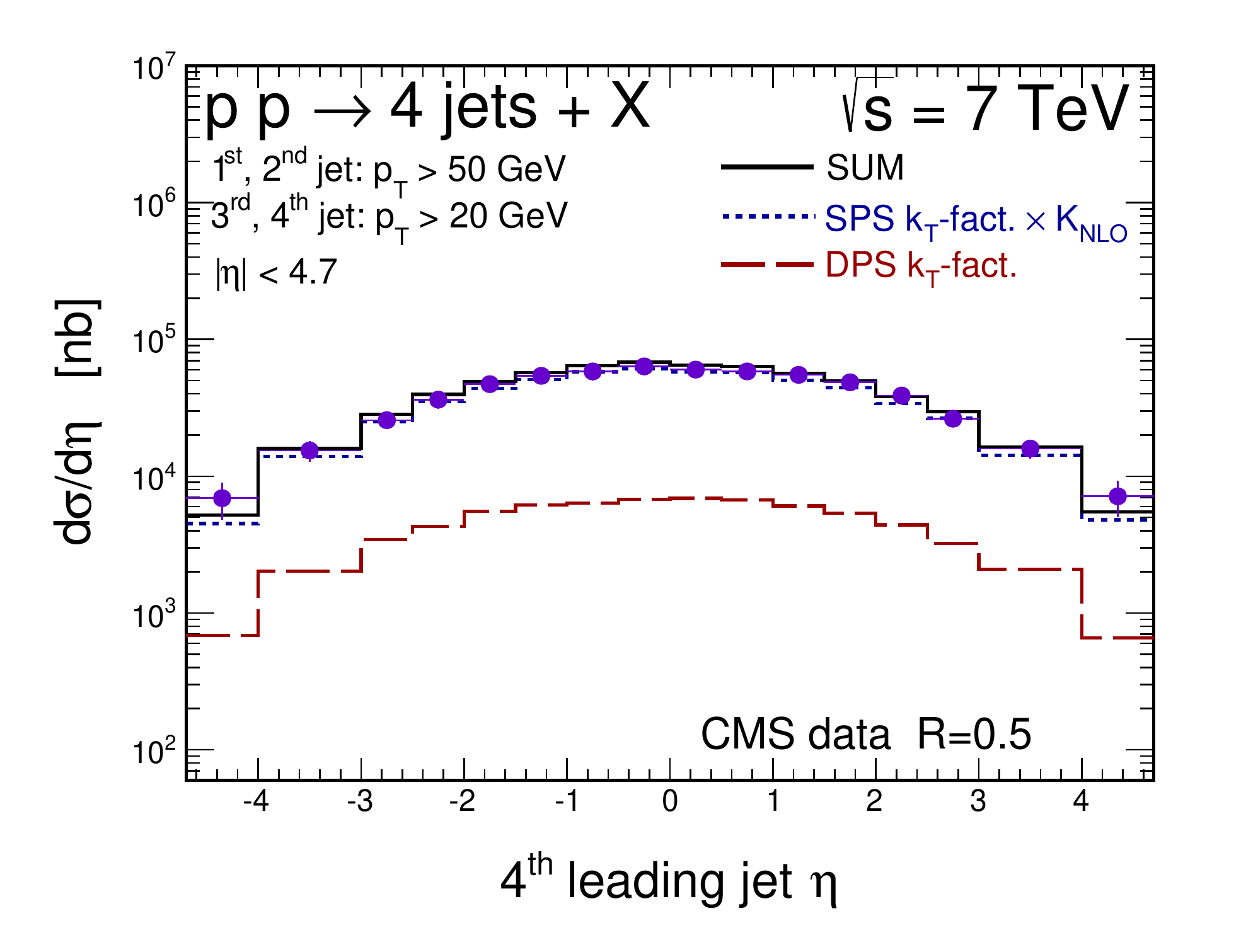}}
\end{minipage}
   \caption{
\small Rapidity distribution of the leading, 2nd, 3rd and 4th jets.
The SPS contribution is shown by the dotted line while the DPS contribution by the dashed line.
 }
 \label{fig:CMS_y_distributions}
\end{figure}

We start our analysis by confronting our approach with the existing data for relatively low cuts on jet transverse momenta.
In this context, the CMS data \cite{Chatrchyan:2013qza} appear to be more suitable than any other available experimental 
analysis of multi-jet production, as they are the only ones featuring sufficiently soft cuts on the transverse momenta 
for DPS to stand out. The cuts on transverse momenta are in this case $|p_T| > 50$ GeV for the two hardest jets 
and $|p_T| > 20$ GeV for the third and fourth ones; the rapidity region is defined by $|\eta| < 4.7$ 
and the constraint on the jet radius parameter is $\Delta R >0.5$.
The situation is shown in Fig.~\ref{fig:CMS_y_distributions}, 
where we plot rapidity distributions for jets ordered by their transverse momenta (leading, 2nd, 3rd, 4th).

The $k_T$-factorization approach includes higher-order corrections 
through the resummation in the PDFs, neglecting fixed order loop effects.
Therefore, we allow for an effective $K$-factor.
From \cite{Bern:2011ep}, the NLO $K$-factors are known to be smaller than unity
for three- and four-jet production in the collinear case with the hard cuts on the transverse momenta 
chosen by the ATLAS collaboration in \cite{Aad:2011tqa} . 
To describe the CMS data, we also need $K$-factors smaller than unity for the SPS contributions, as expected.
Concerning the DPS contribution, instead, we do not include $K$ factors and the motivation is as follows. 
The theoretical $K$-factor for the 2-jet inclusive cross section in the collinear case and for the same cuts as above 
is known to be $~1.18$ or $~1.25$, depending on whether one includes or not non perturbative hadronization 
effects on top of the NLO calculation. 
But, contrary to the three- and four-jet cases, the NLO predictions for the inclusive cross section 
is further away from the measured value than the LO one \cite{Bern:2011ep}. 
This is due to a phase space effect which is specific to 2-jet production at fixed perturbative order
and affects primarily the lowest $p_T$ bins, as first discussed in \cite{Frixione:1997ks} and 
remarked, from another point of view, in \cite{Kutak:2016mik} 
(for another recent discussion of such effect in two jet production in the context of DIS, see \cite{Currie:2016ytq}).
The resulting overestimation of the cross section is the reason why
the theoretical 2-jet $K$-factors would lead to an overestimation of DPS.

We use $\sigma_{eff}$ = 15 mb in the pocket-formula (\ref{pocket}) to calculate the DPS contribution.
This is a typical value known from the world systematics \cite{Proceedings:2016tff}.
However, in the present study we consider larger energies and we explore a slightly different region, 
and such a value does not need to be universal. 
Larger values of $\sigma_{eff}$ were obtained recently, for example, 
for $D$ meson production when including $g \to D$ fragmentation \cite{Maciula:2016wci}.

In the following, we will propose a set of observables that we find particularly convenient to
identify DPS effects in four-jet production, both for symmetric and asymmetric cuts.

\begin{figure}[!h]
{\includegraphics[width=0.6\textwidth]{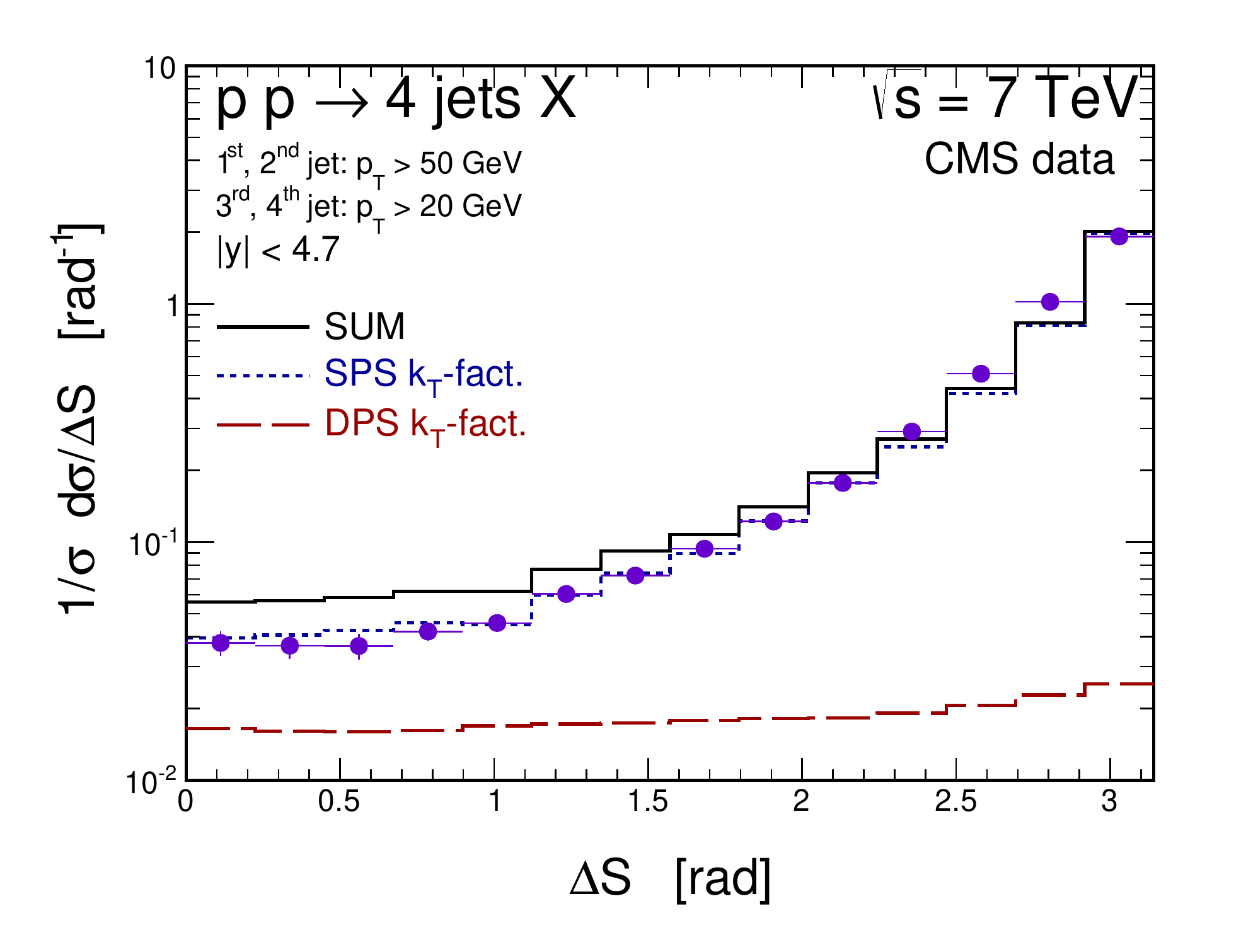}}
\caption{ \small Distribution in the $\Delta S$ variable. The SPS contribution is shown by the dotted line while the DPS contribution by the dashed line. }
 \label{fig:CMS_DeltaS_distribution}
\end{figure}

Some comments are in order concerning Fig. \ref{fig:CMS_DeltaS_distribution},
showing the plot of distribution in the variable which was proposed as a potential smoking gun 
for DPS in four-jet production \cite{Chatrchyan:2013qza} which is the azimuthal 
angle separation between the hardest and softest pair of jets.
This variable is defined as a ratio between a differential and a total cross section, 
which makes it insensitive to possibly constant K-factors from higher order corrections;
only a phase-space dependence of the K-factors could have an impact on this distribution.
Setting this hypothesis aside for the moment, as one can see, the SPS contribution 
computed with our $k_T$-factorization approach describes the data pretty well within uncertainties, 
except for two of the highest bins. The situation in the highest bins does not seem significantly improved 
by the DPS contribution, which otherwise leads to overestimation of the data in the lower bins.

Considering the mentioned proviso on phase space dependence, our conclusion is that it is best to propose other variables
which, on the ground of the theoretical calculation, seem potentially useful in discriminating more clearly between
SPS and DPS in four-jet production.

\subsection{Symmetric cuts}

\begin{figure}[!h]
\begin{minipage}{0.47\textwidth}
 \centerline{\includegraphics[width=1.0\textwidth]{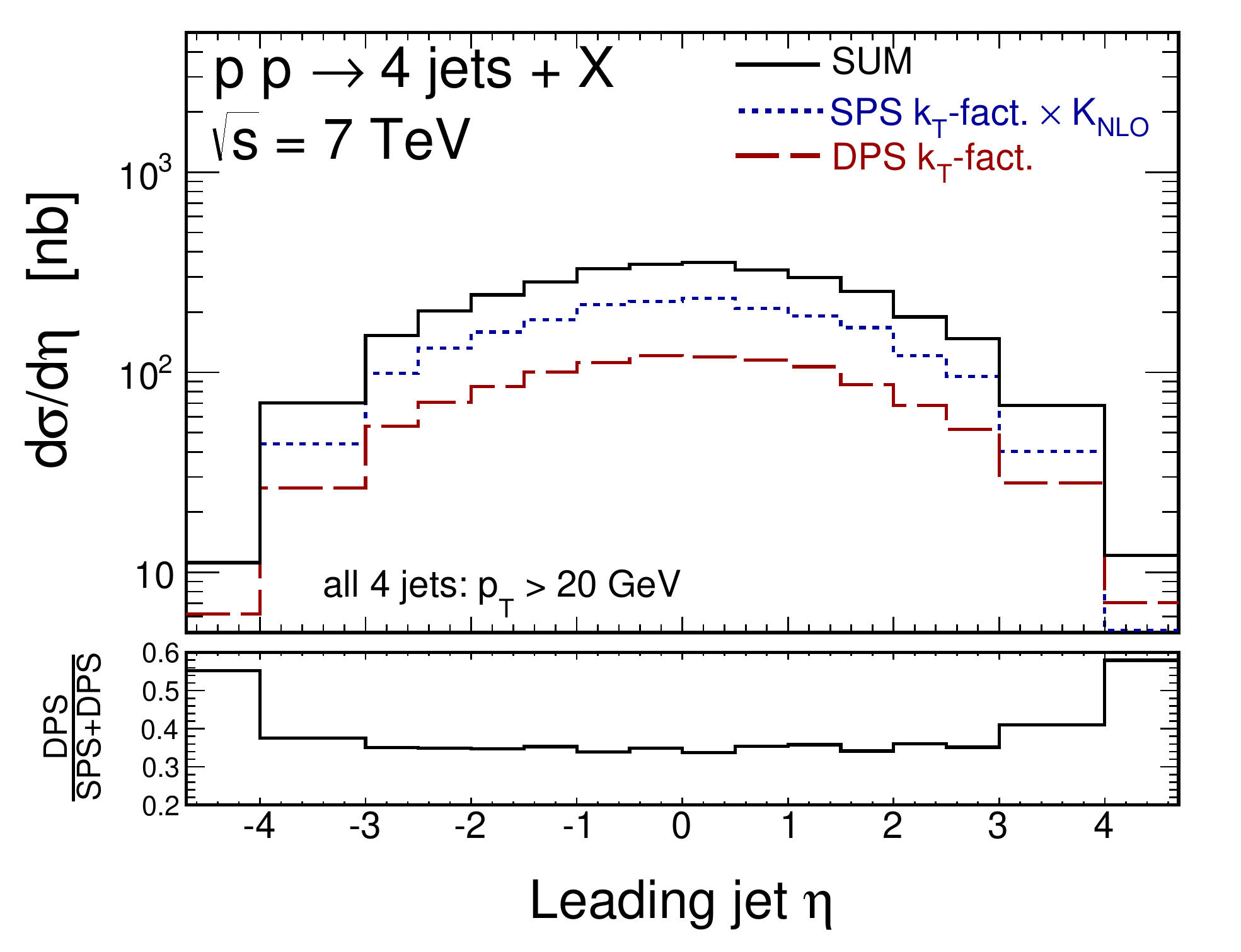}}
\end{minipage}
\hspace{0.5cm}
\begin{minipage}{0.47\textwidth}
 \centerline{\includegraphics[width=1.0\textwidth]{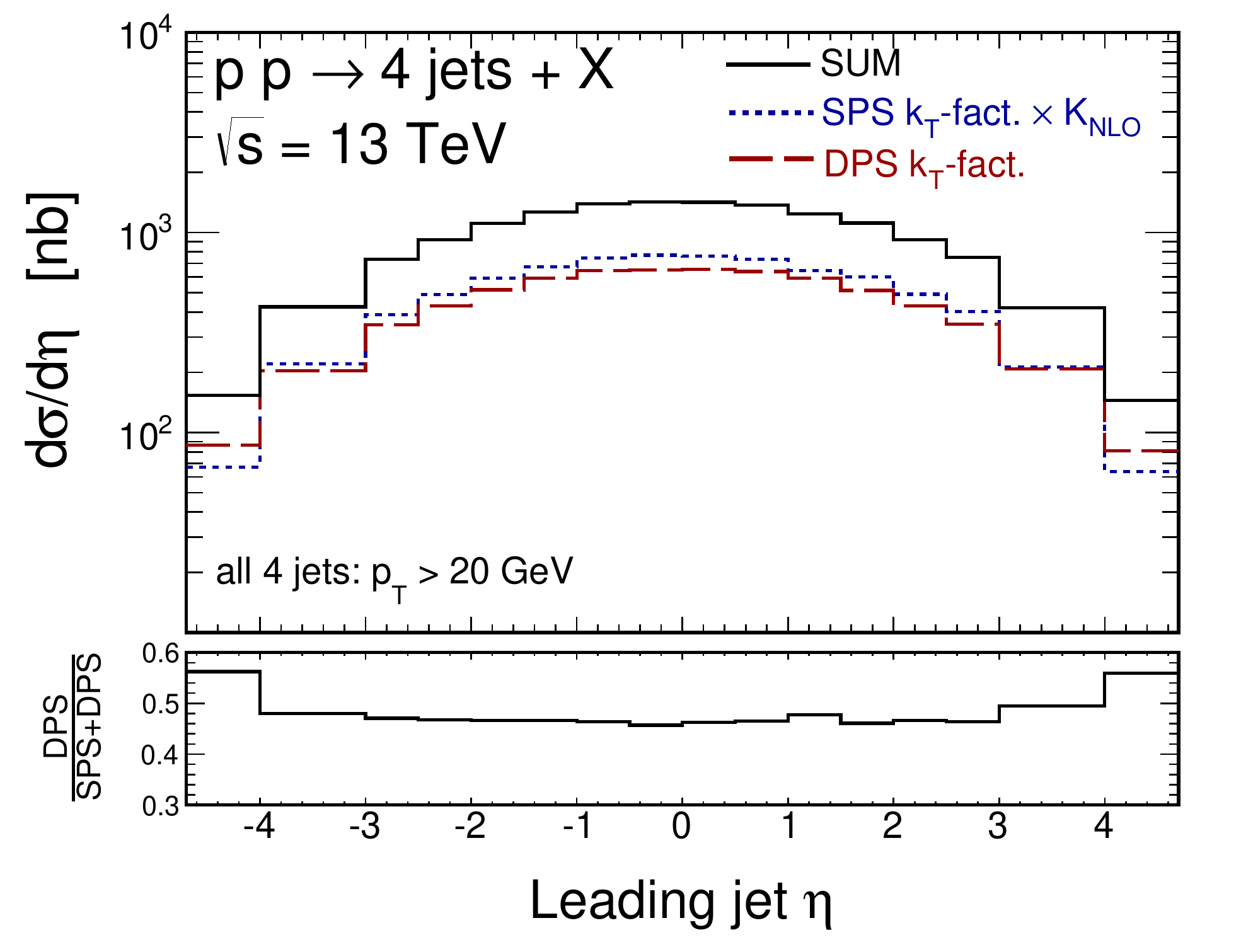}}
\end{minipage}\\
\begin{minipage}{0.47\textwidth}
 \centerline{\includegraphics[width=1.0\textwidth]{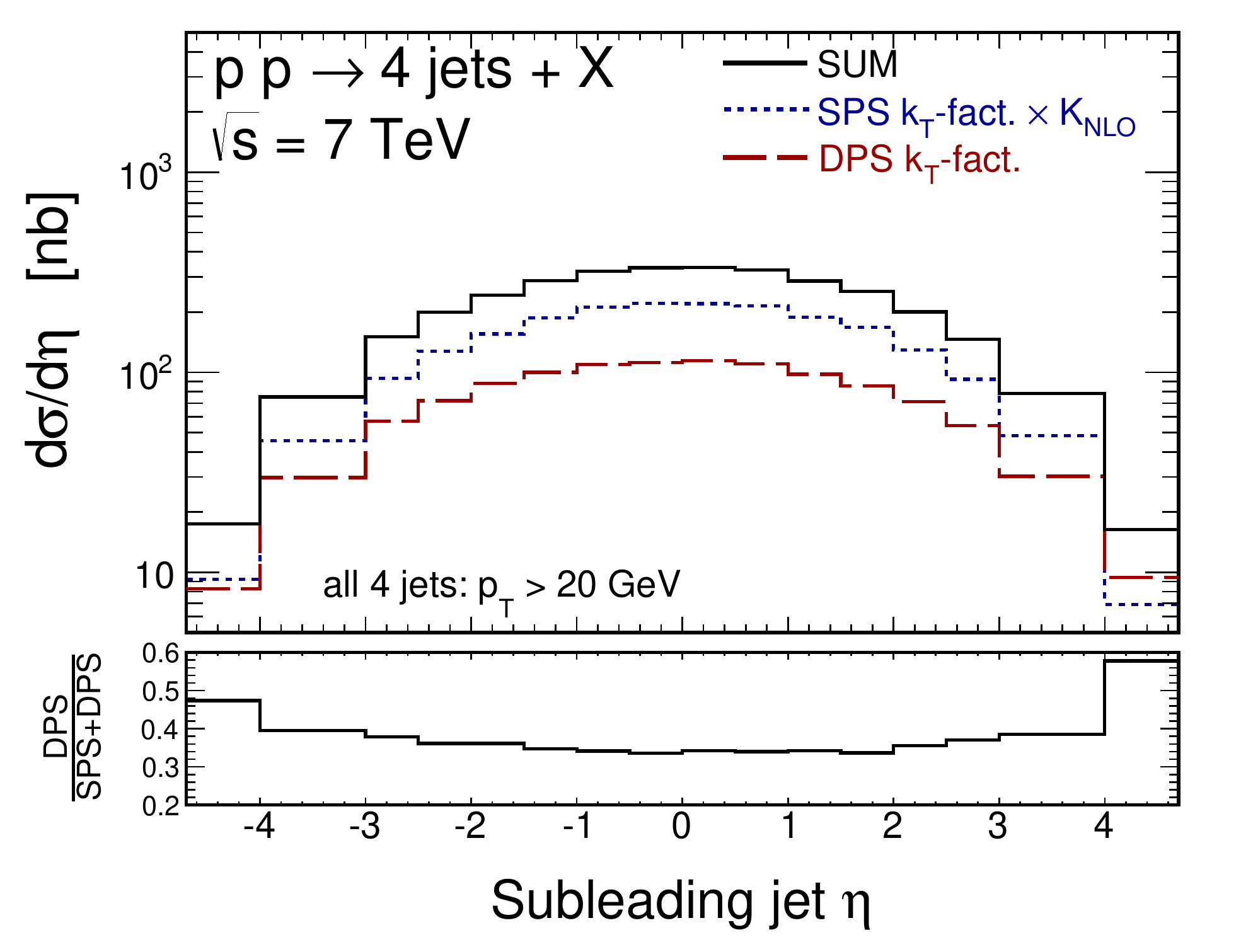}}
\end{minipage}
\hspace{0.5cm}
\begin{minipage}{0.47\textwidth}
 \centerline{\includegraphics[width=1.0\textwidth]{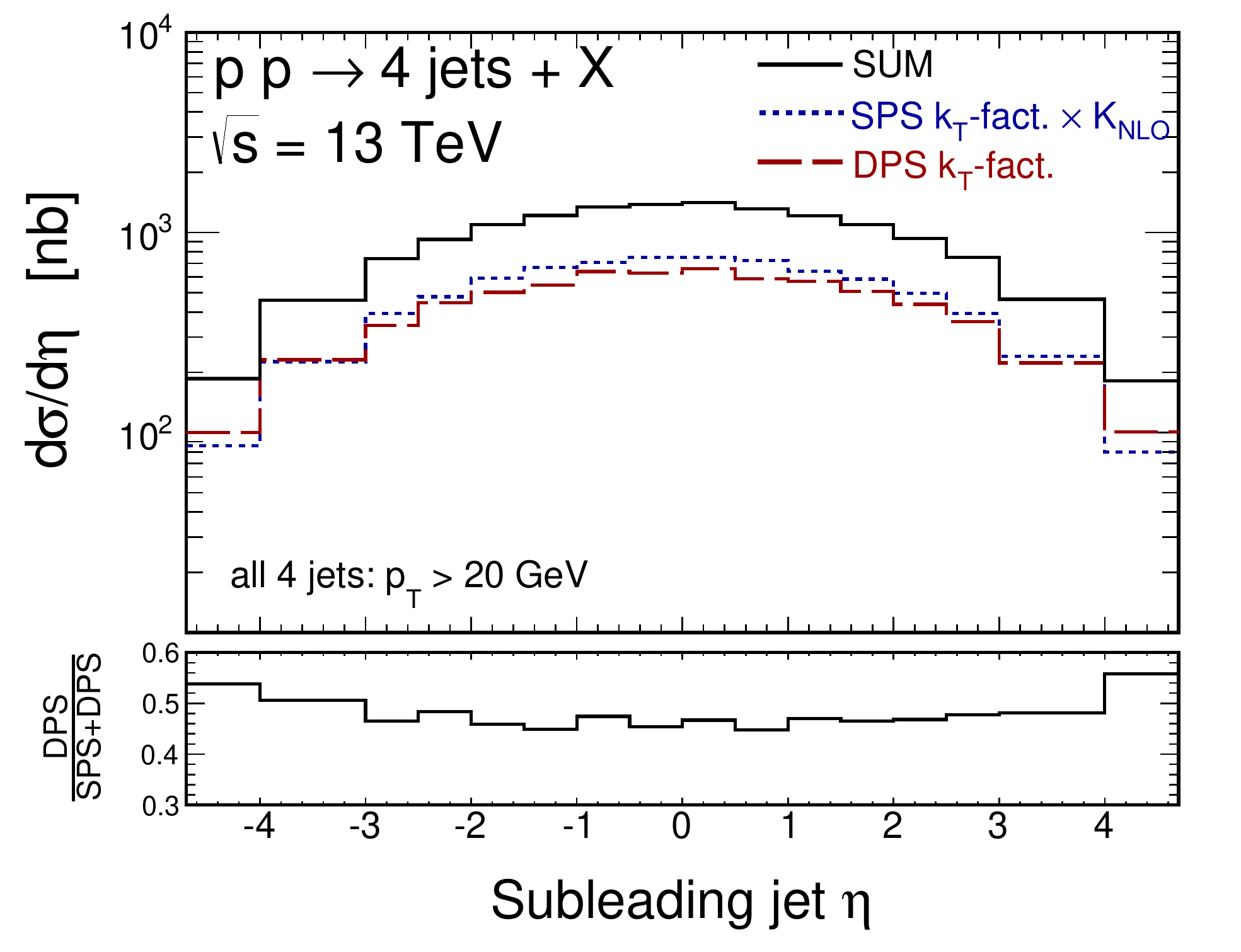}}
\end{minipage}
   \caption{
\small Rapidity distribution of leading and subleading jets for $\sqrt{s}$ = 7 TeV
(left column) and $\sqrt{s}$ = 13 TeV (right column) for
the symmetric cuts. The SPS contribution
is shown by the dotted line while the DPS contribution by the dashed line.
The relative contribution of DPS is shown in the extra lower panels.
 }
 \label{fig:dsig_dy_symmetric_20GeV}
\end{figure}

In this section we introduce our proposed optimal observables for the study of DPS.
We start with a completely symmetric cuts scenario, $p_T > 20$ GeV for all the four leading jets,
moving on to the asymmetric case in the following section. In both this and the following 
section the cuts on rapidity and jet radius parameter stay the same as for the CMS case.
In Fig.~\ref{fig:dsig_dy_symmetric_20GeV} we show our predictions
for the rapidity distributions. In contrast to the previous case (Fig.~\ref{fig:CMS_y_distributions}), which
featured a harder cut on the two hardest jets, the shapes of the SPS and DPS rapidity distributions 
are rather similar. There is only a small relative enhancement of the DPS contribution for larger jet rapidities $|\eta|$.
This is also different than the result obtained in the leading-order collinear approach \cite{Maciula:2015vza}.

Elaborating on the results of \cite{Maciula:2014pla}, 
it was shown in Ref.~\cite{Maciula:2015vza} in a collinear approach
that two more observables are potentially
useful to nail down DPS, namely the maximum rapidity distance 
\begin{equation}
\Delta \text{Y} \equiv max_{\substack{i,j \in\{1,2,3,4\}\\i \neq j }} |\eta_i-\eta_j |
\end{equation}
and the azimuthal correlations between the jets which are most remote in rapidity
\begin{equation}
\varphi_{jj} \equiv | \varphi_i -\varphi_j |  \, , \quad \text{for}  \quad |\eta_i - \eta_j | = \Delta \text{Y} \, .
\end{equation}

\begin{figure}[!h]
\begin{minipage}{0.47\textwidth}
 \centerline{\includegraphics[width=1.0\textwidth]{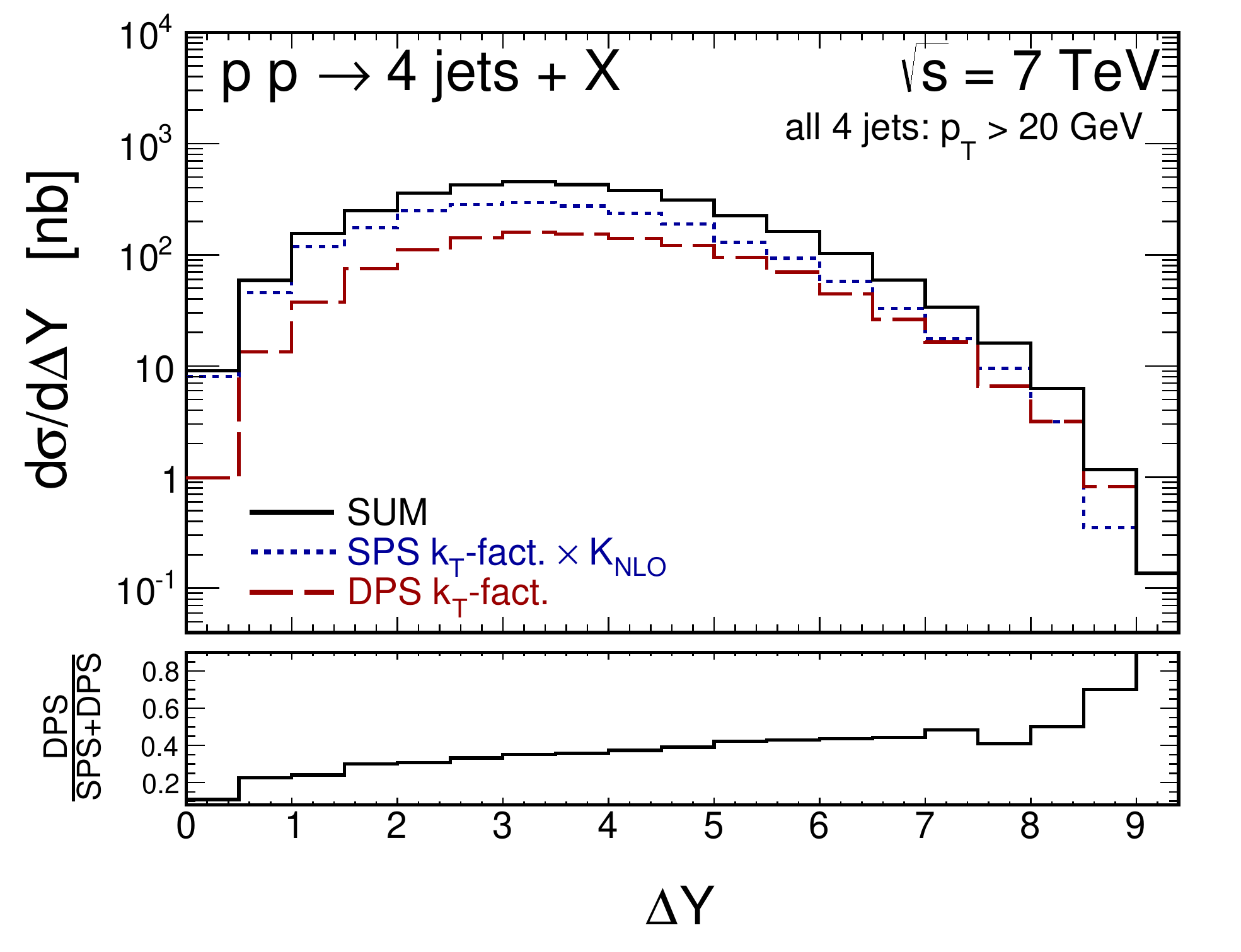}}
\end{minipage}
\hspace{0.5cm}
\begin{minipage}{0.47\textwidth}
 \centerline{\includegraphics[width=1.0\textwidth]{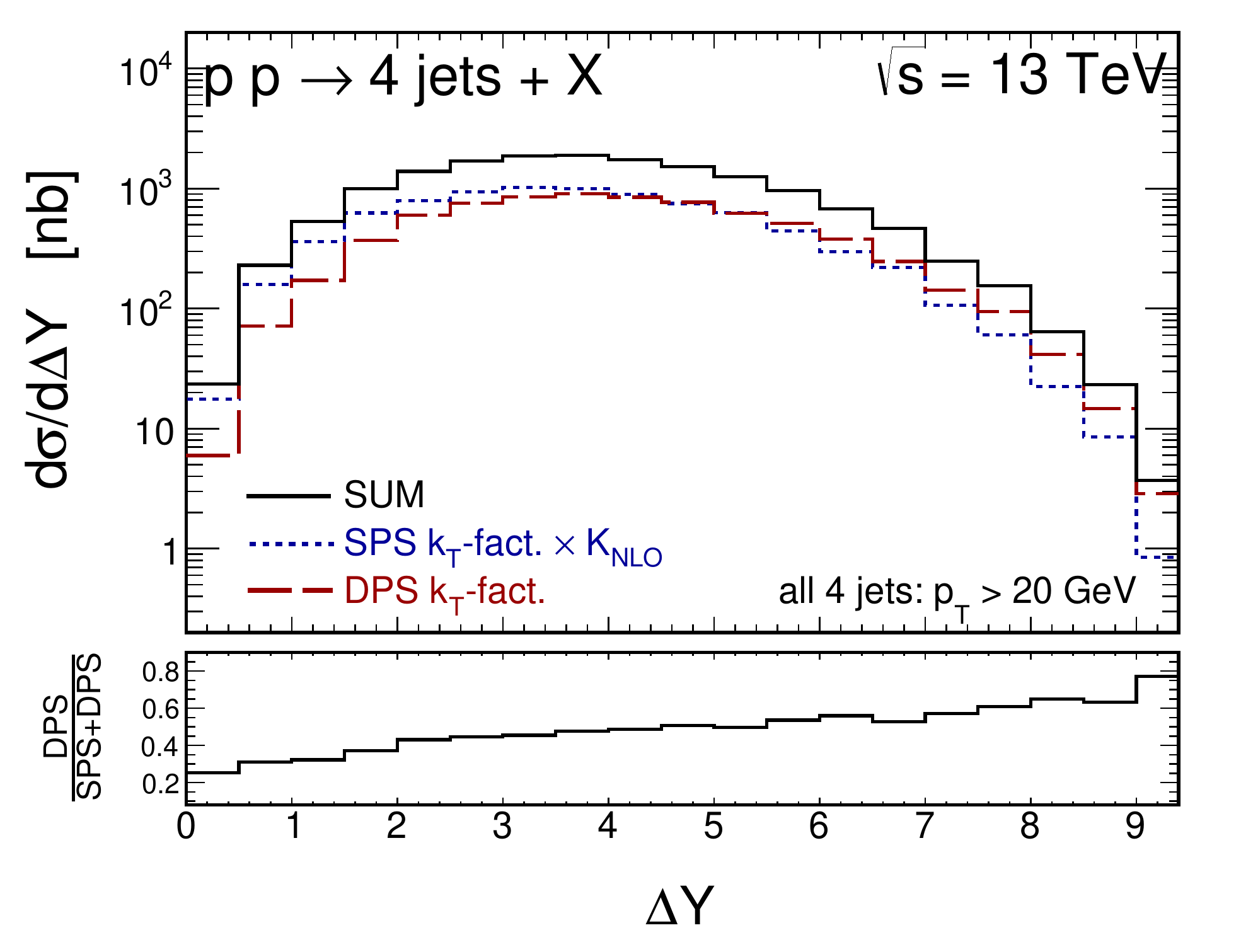}}
\end{minipage}
   \caption{
\small Distribution in rapidity distance between the most remote jets for 
the symmetric cut with $p_T >$ 20 GeV 
for $\sqrt{s}$ = 7 TeV (left) and $\sqrt{s}$ = 13 TeV (right).
The SPS contribution is shown by the dotted line while 
the DPS contribution by the dashed line.
The relative contribution of DPS is shown in the extra lower panels.
 }
 \label{fig:dsig_dydiff}
\end{figure}

One can see in Fig.~\ref{fig:dsig_dydiff} that the relative DPS contribution gradually increases 
with $\Delta \text{Y}$ which, for the CMS collaboration, can be as large as 9.4.
A potential failure of the SPS contribution to describe such a plot
would therefore be a signal of the presence of a sizable DPS contribution.

\begin{figure}[!h]
\begin{minipage}{0.47\textwidth}
 \centerline{\includegraphics[width=1.0\textwidth]{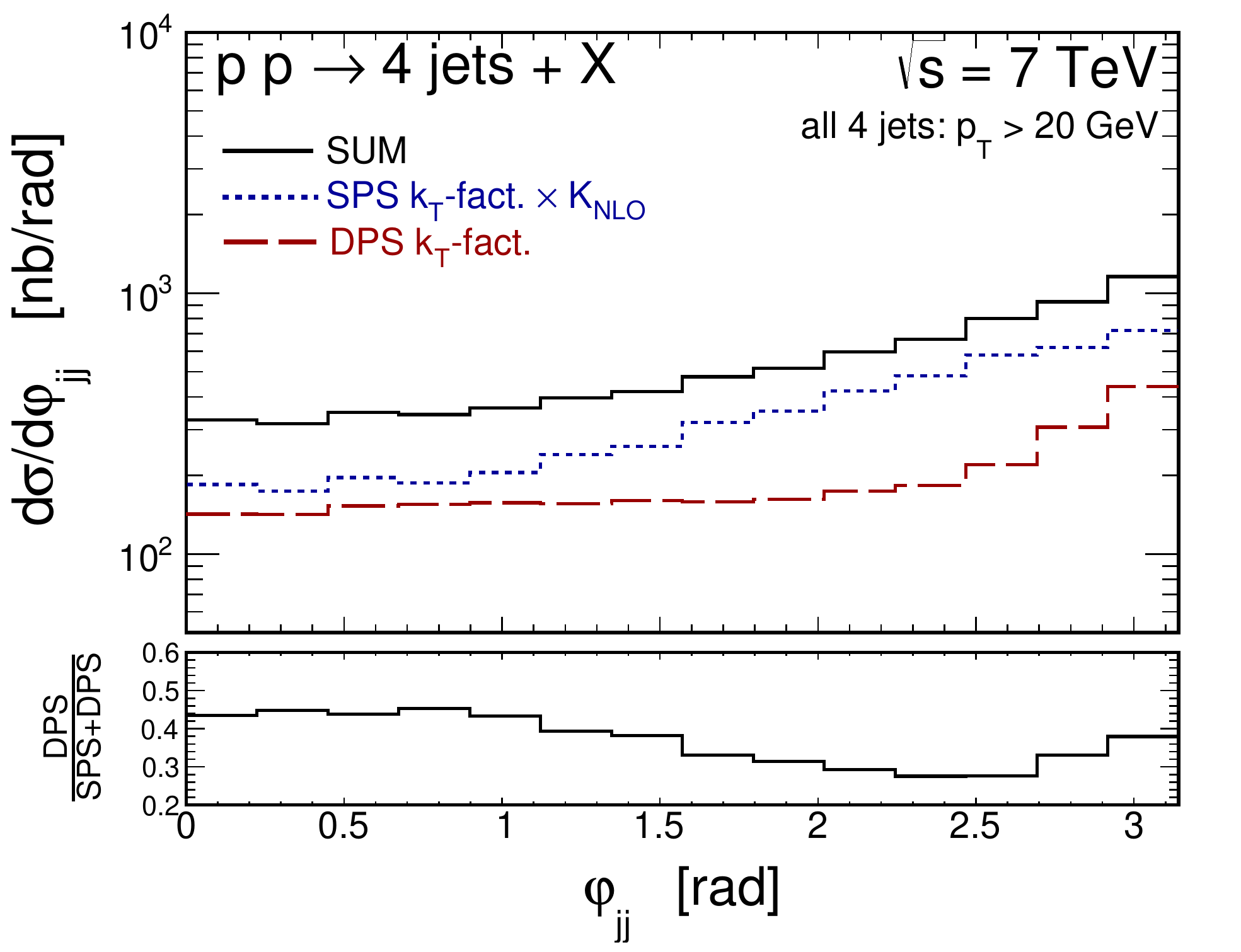}}
\end{minipage}
\hspace{0.5cm}
\begin{minipage}{0.47\textwidth}
 \centerline{\includegraphics[width=1.0\textwidth]{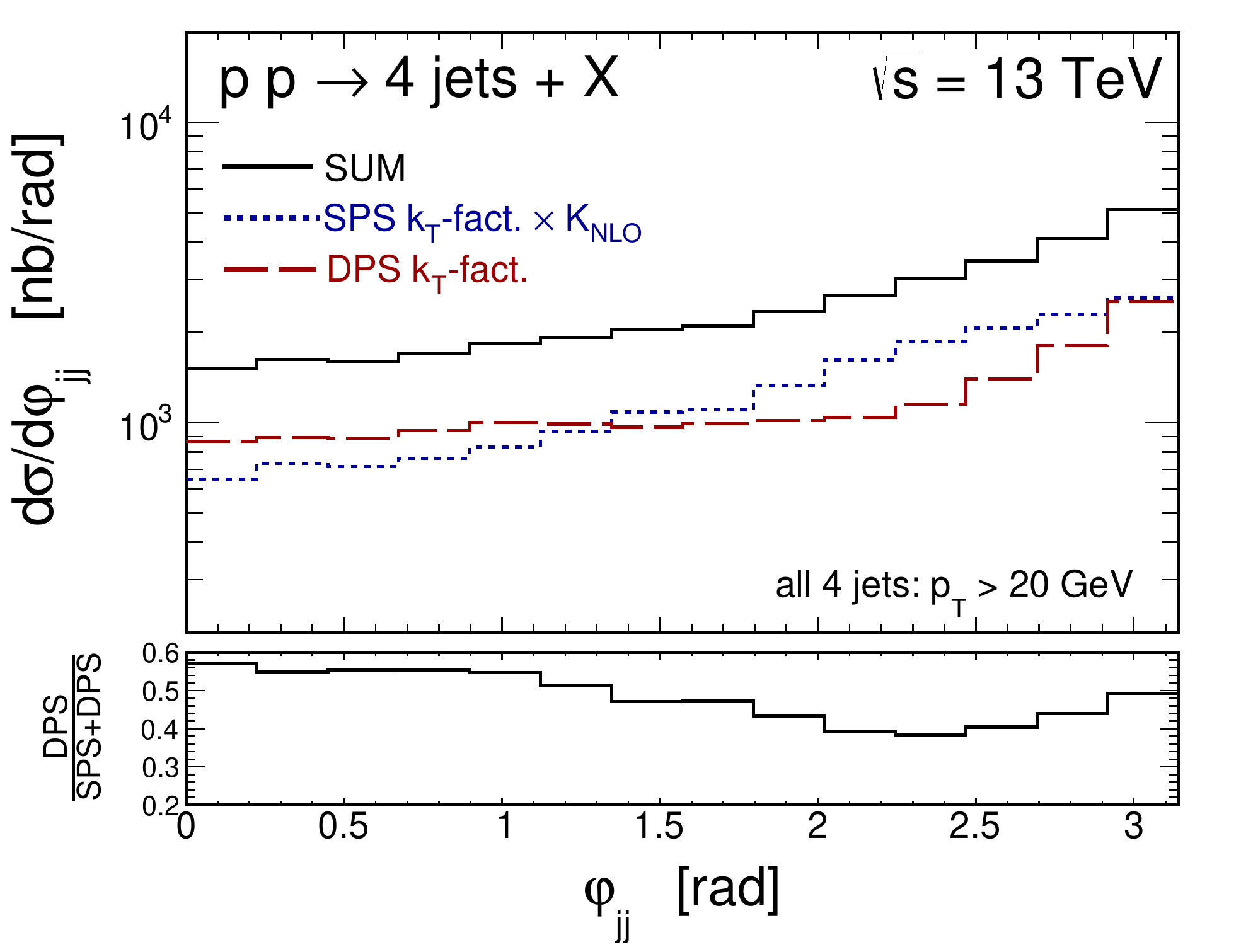}}
\end{minipage}
   \caption{
\small Distribution in relative azimuthal angle between the most remote jets for 
the symmetric cut with $p_T >$ 20 GeV 
for $\sqrt{s}$ = 7 TeV (left) and $\sqrt{s}$ = 13 TeV (right).
The SPS contribution is shown by the dotted line while 
the DPS contribution by the dashed line.
The relative contribution of DPS is shown in the extra lower panels.
 }
 \label{fig:dsig_dphijj}
\end{figure}

Fig.~\ref{fig:dsig_dphijj} depicts azimuthal correlations between the jets most remote in rapidity. 
While at $\sqrt{s}$ = 7 TeV the SPS contribution is always larger than the DPS contribution, 
at $\sqrt{s}$ = 13 TeV the DPS contribution dominates over the SPS contribution for $\varphi_{jj} < \pi/2$. 
The relative DPS contribution is shown again in the lower extra panels.

\begin{figure}[!h]
\begin{minipage}{0.47\textwidth}
 \centerline{\includegraphics[width=1.0\textwidth]{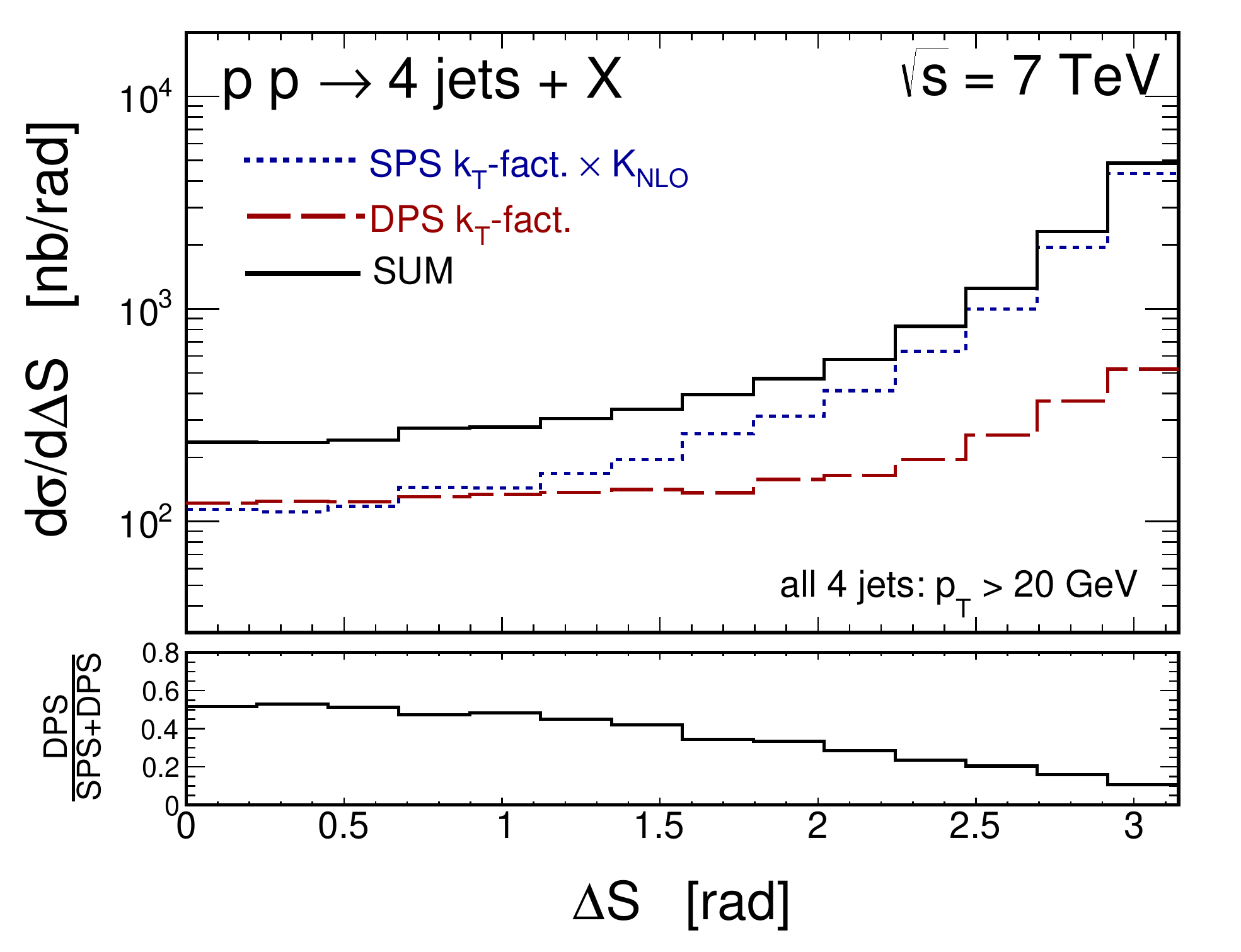}}
\end{minipage}
\hspace{0.5cm}
\begin{minipage}{0.47\textwidth}
 \centerline{\includegraphics[width=1.0\textwidth]{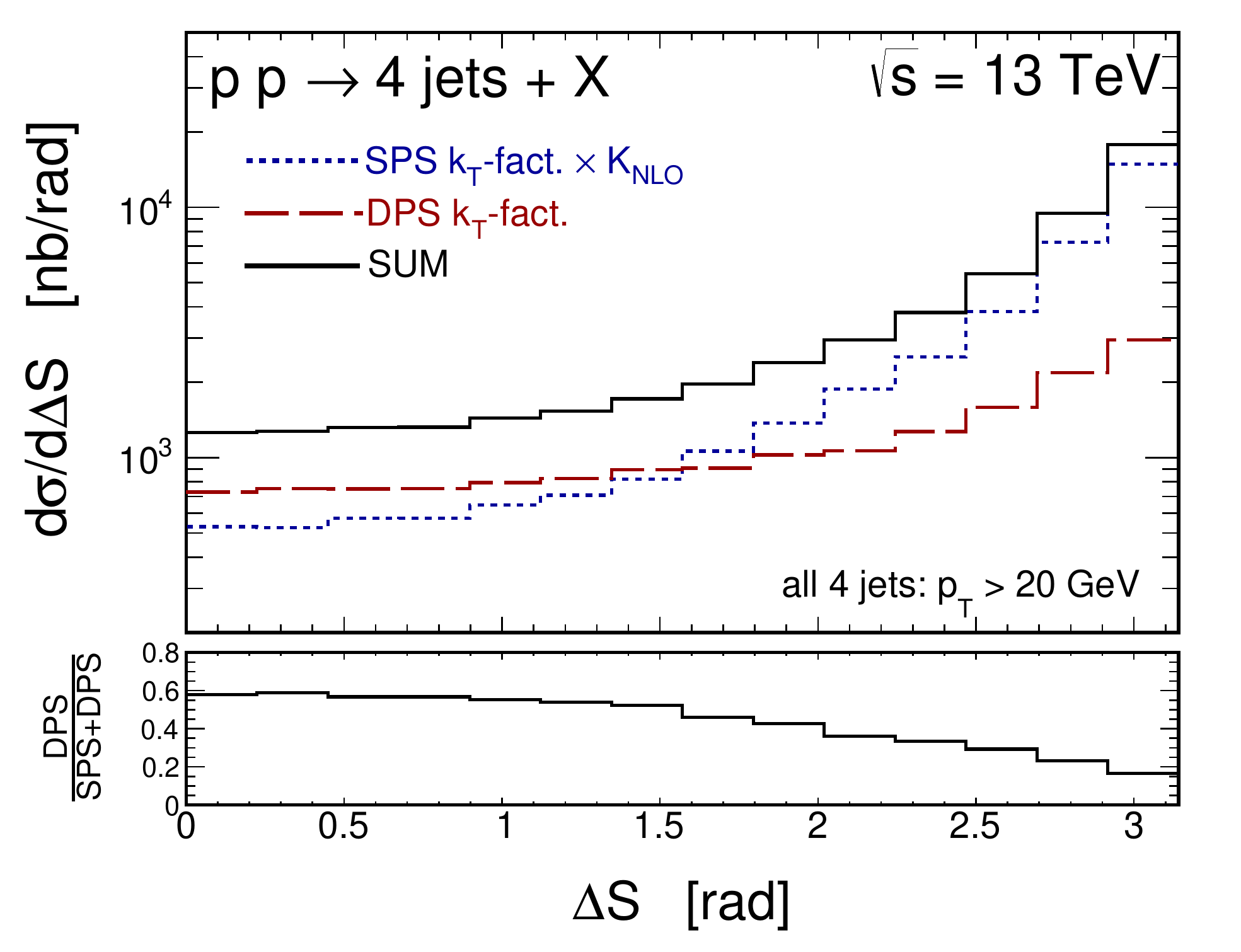}}
\end{minipage}
   \caption{
\small Distribution in $\Delta S$ for 
the symmetric cut with $p_T >$ 20 GeV 
for $\sqrt{s}$ = 7 TeV (left) and $\sqrt{s}$ = 13 TeV (right).
The SPS contribution is shown by the dotted line while 
the DPS contribution by the dashed line.
The relative contribution of DPS is shown in the extra lower panels.
 }
 \label{fig:dsig_dDeltaS}
\end{figure}

In Fig.~\ref{fig:dsig_dDeltaS} we show distribution in the $\Delta S$ variable already discussed for the CMS cuts (see Fig.~\ref{fig:CMS_DeltaS_distribution}). Here the relative contribution of the DPS is bigger than for the CMS experiment.
For $\sqrt{s} = 13$ TeV the DPS component wins with the SPS one for $\Delta S < \frac{\pi}{2}$. 

\begin{figure}[!h]
\begin{minipage}{0.47\textwidth}
 \centerline{\includegraphics[width=1.0\textwidth]{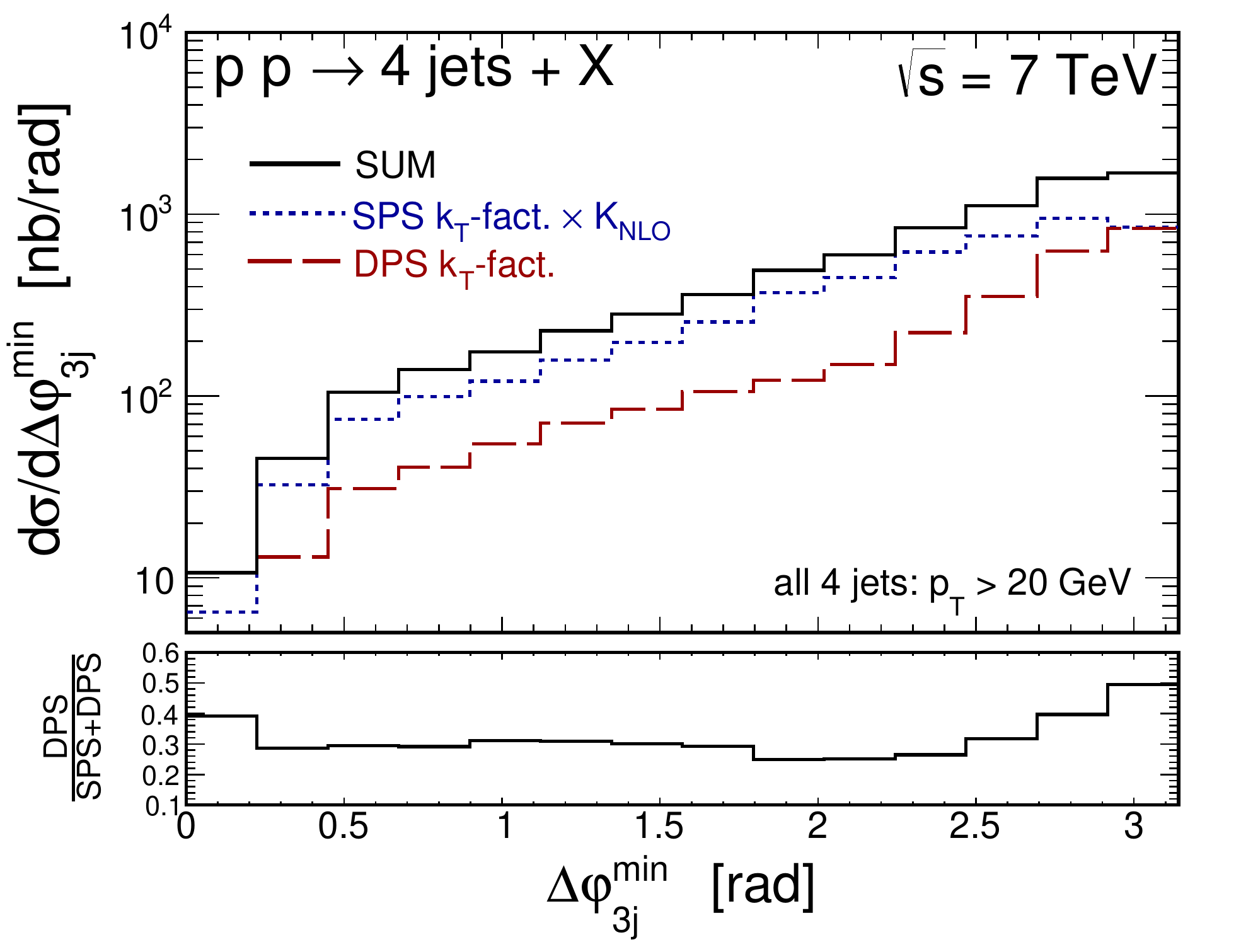}}
\end{minipage}
\hspace{0.5cm}
\begin{minipage}{0.47\textwidth}
 \centerline{\includegraphics[width=1.0\textwidth]{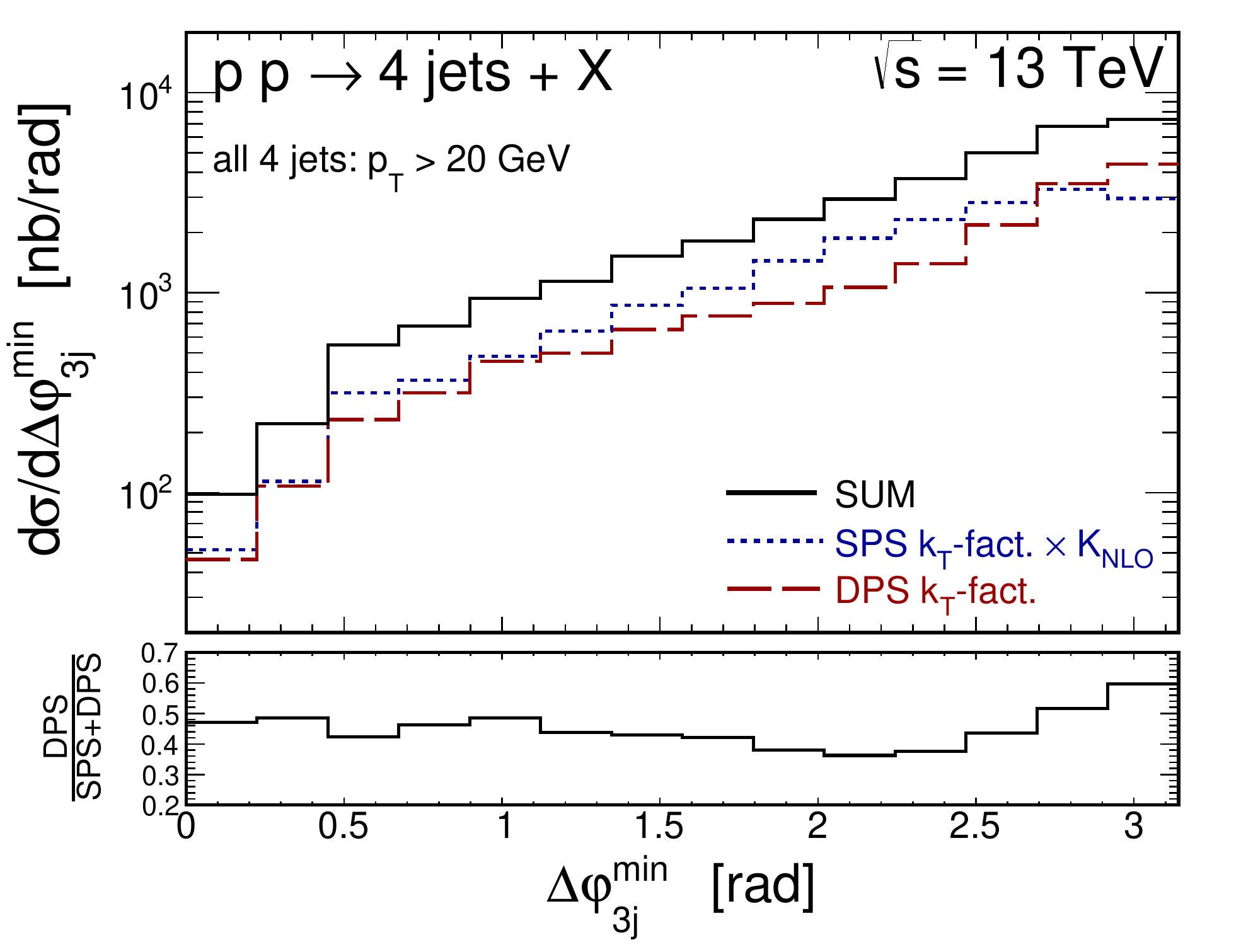}}
\end{minipage}
   \caption{
\small Distribution in $\Delta \varphi_{3j}^{min}$ angle for 
the symmetric cut with $p_T >$ 20 GeV 
for $\sqrt{s}$ = 7 TeV (left) and $\sqrt{s}$ = 13 TeV (right).
The SPS contribution is shown by the dotted line while 
the DPS contribution by the dashed line.
The relative contribution of DPS is shown in the extra lower panels.
 }
 \label{fig:dsig_dphi_3j}
\end{figure}

We also find that another variable, introduced in the high transverse momenta analysis 
of 4 jets production presented in Ref.~\cite{Aad:2015nda}, can be very interesting for 
the scrutiny of DPS effects. It is defined as follows
\begin{equation}
\Delta \varphi_{3j}^{min} \equiv min_{\substack{i,j,k \in\{1,2,3,4\}\\i 
\neq j \neq k}}\left(|\varphi_i - \varphi_j|+| \varphi_j - \varphi_k|\right) \, .
\label{DeltaPhiMin}
\end{equation}
As three out of four azimuthal angles are always entering in (\ref{DeltaPhiMin}), configurations
featuring one jet recoiling against the other three are necessarily characterised by lower
values of $\Delta \varphi_{3j}^{min}$ with respect to the two-against-two topology;
the minimum, in fact, will be obtained in the first case for $i,j,k$ 
denoting the three jets in the same half hemisphere, 
whereas such a situation is not possible for the second configuration.
Obviously, the first case would be allowed only by SPS in a collinear tree-level framework,
whereas the second should be enhanced by DPS. 
In the $k_T$-factorization approach, 
this situation is smeared out by the presence of transverse momenta of the initial state partons. 
For our TMDs, the corresponding distributions are shown in Fig.~\ref{fig:dsig_dphi_3j}.
In contrast to the naive expectations, similar shapes are obtained for DPS and SPS contributions.

\subsection{Asymmetric cuts}

So far we have addressed the problem of identifying DPS with low and 
completely symmetric cuts on the transverse momenta.
Nevertheless, as already remarked above, it was also pointed out in \cite{Kutak:2016mik}, that the two-jet production mechanism
which accounts for DPS is affected by a severe underestimation of the cross section when higher order effects are included.
It is thus desirable, in order to get rid of this phase-space effect when looking for DPS, 
to employ asymmetric cuts on the jets transverse momenta, especially when considering $p_T$ distributions.

\begin{figure}[!h]
\begin{minipage}{0.47\textwidth}
 \centerline{\includegraphics[width=1.0\textwidth]{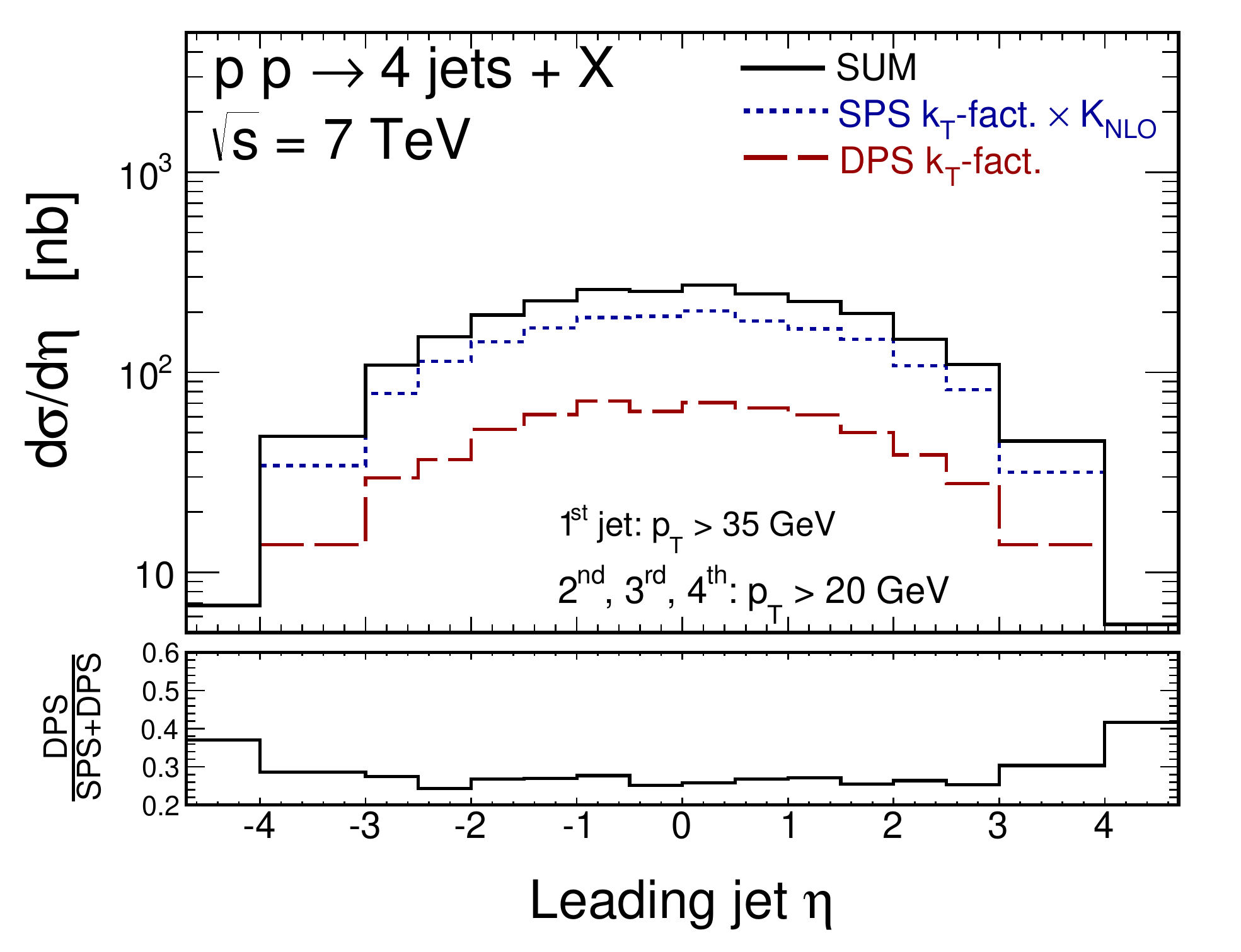}}
\end{minipage}
\hspace{0.5cm}
\begin{minipage}{0.47\textwidth}
 \centerline{\includegraphics[width=1.0\textwidth]{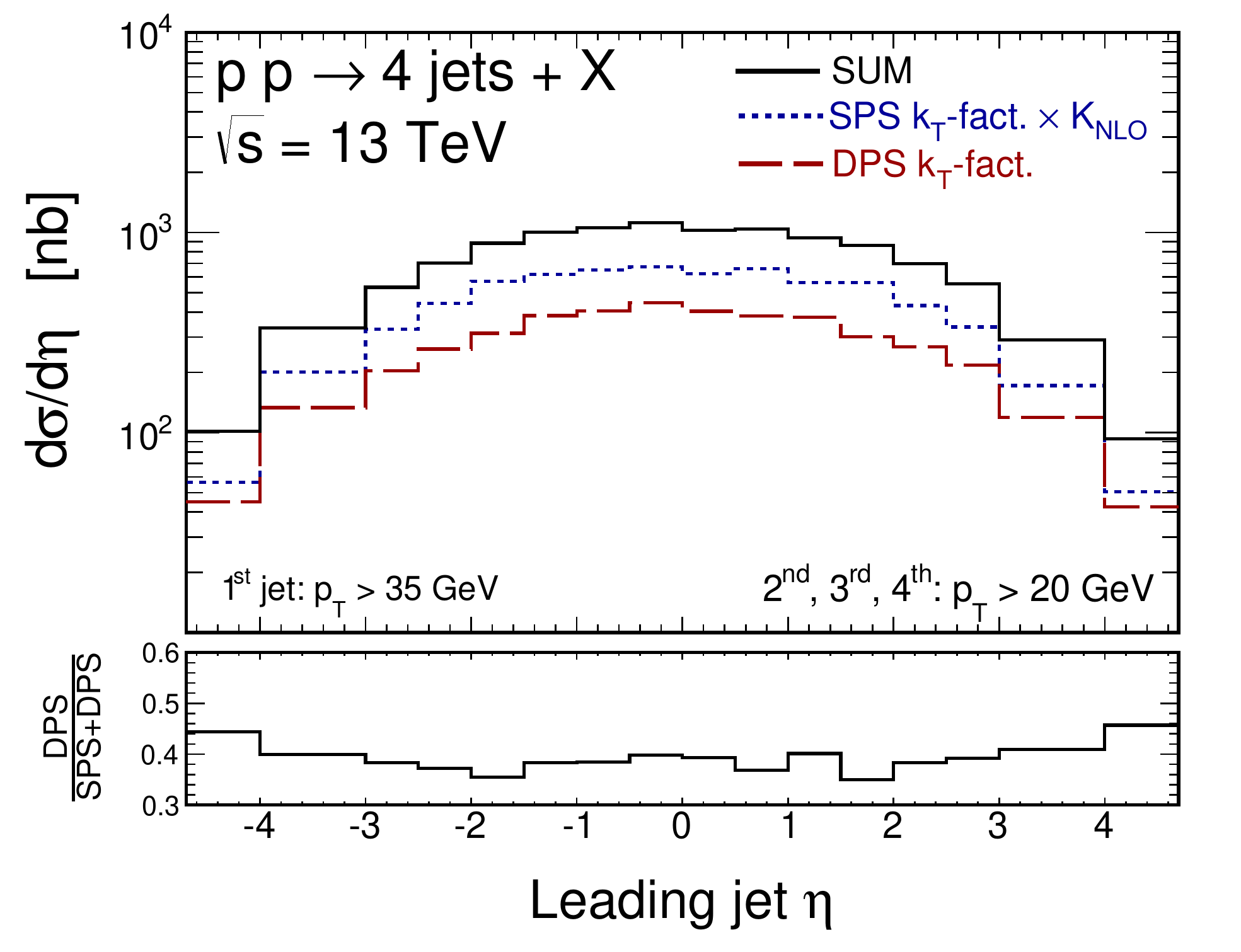}}
\end{minipage}\\
\begin{minipage}{0.47\textwidth}
 \centerline{\includegraphics[width=1.0\textwidth]{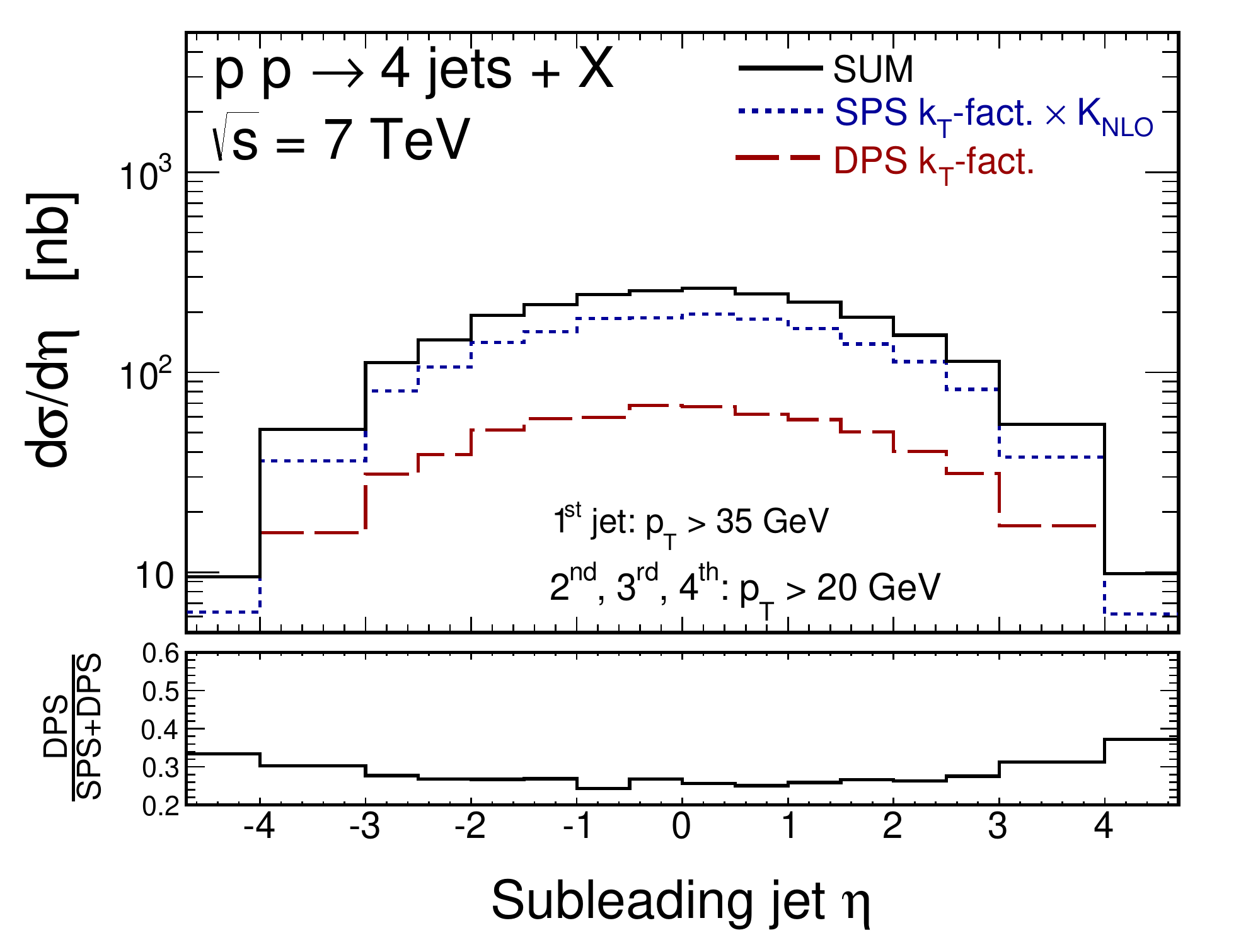}}
\end{minipage}
\hspace{0.5cm}
\begin{minipage}{0.47\textwidth}
 \centerline{\includegraphics[width=1.0\textwidth]{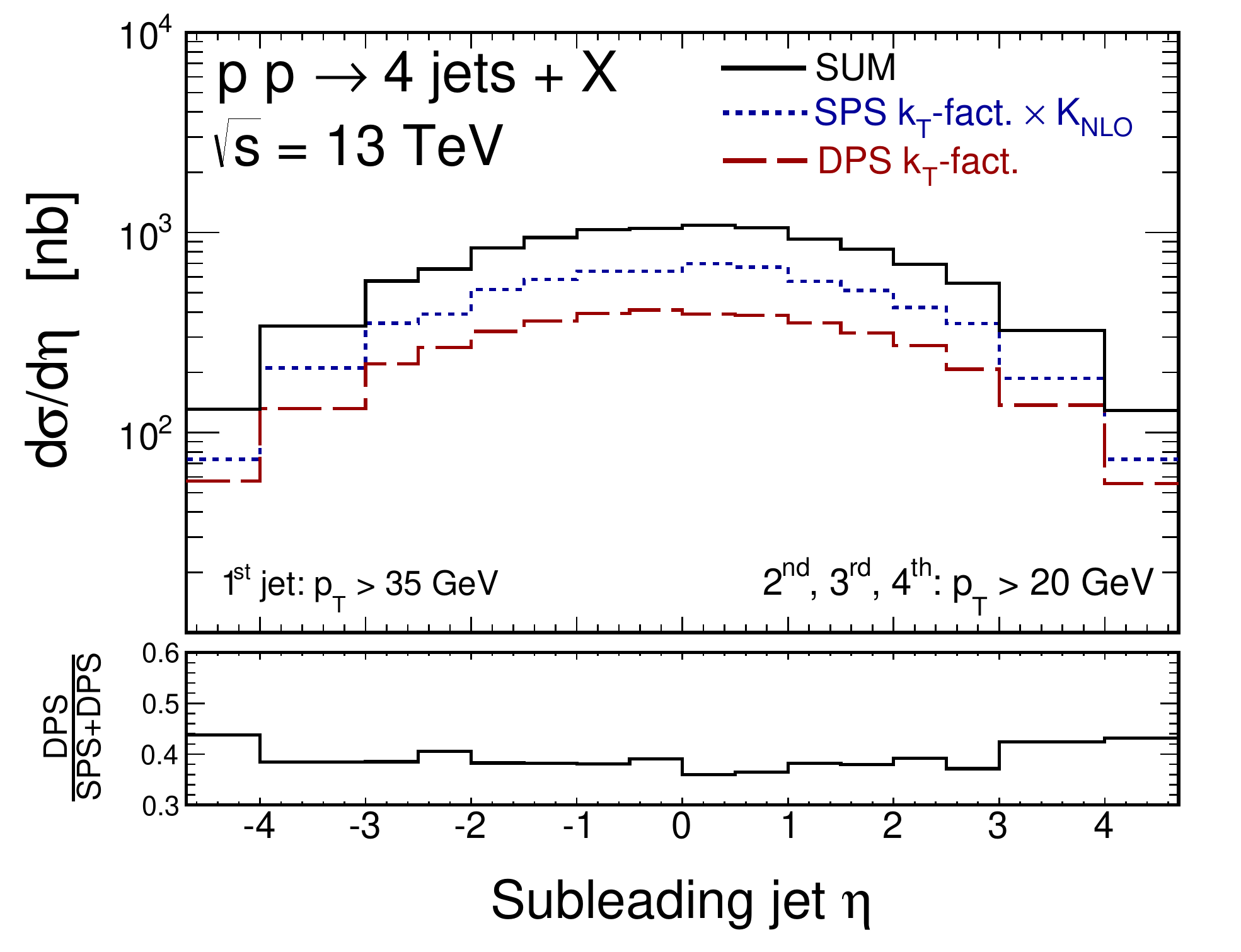}}
\end{minipage}
   \caption{
\small Rapidity distribution of leading and subleading jets for $\sqrt{s}$ = 7 TeV
(left column) and $\sqrt{s}$ = 13 TeV (right column) for
the asymmetric cuts. The SPS contribution
is shown by the dotted line while the DPS contribution by the dashed line.
The relative contribution of DPS is shown in the extra lower panels.
 }
 \label{fig:dsig_dy_asymmetric}
\end{figure}

\begin{figure}[h]
\begin{minipage}{0.47\textwidth}
 \centerline{\includegraphics[width=1.0\textwidth]{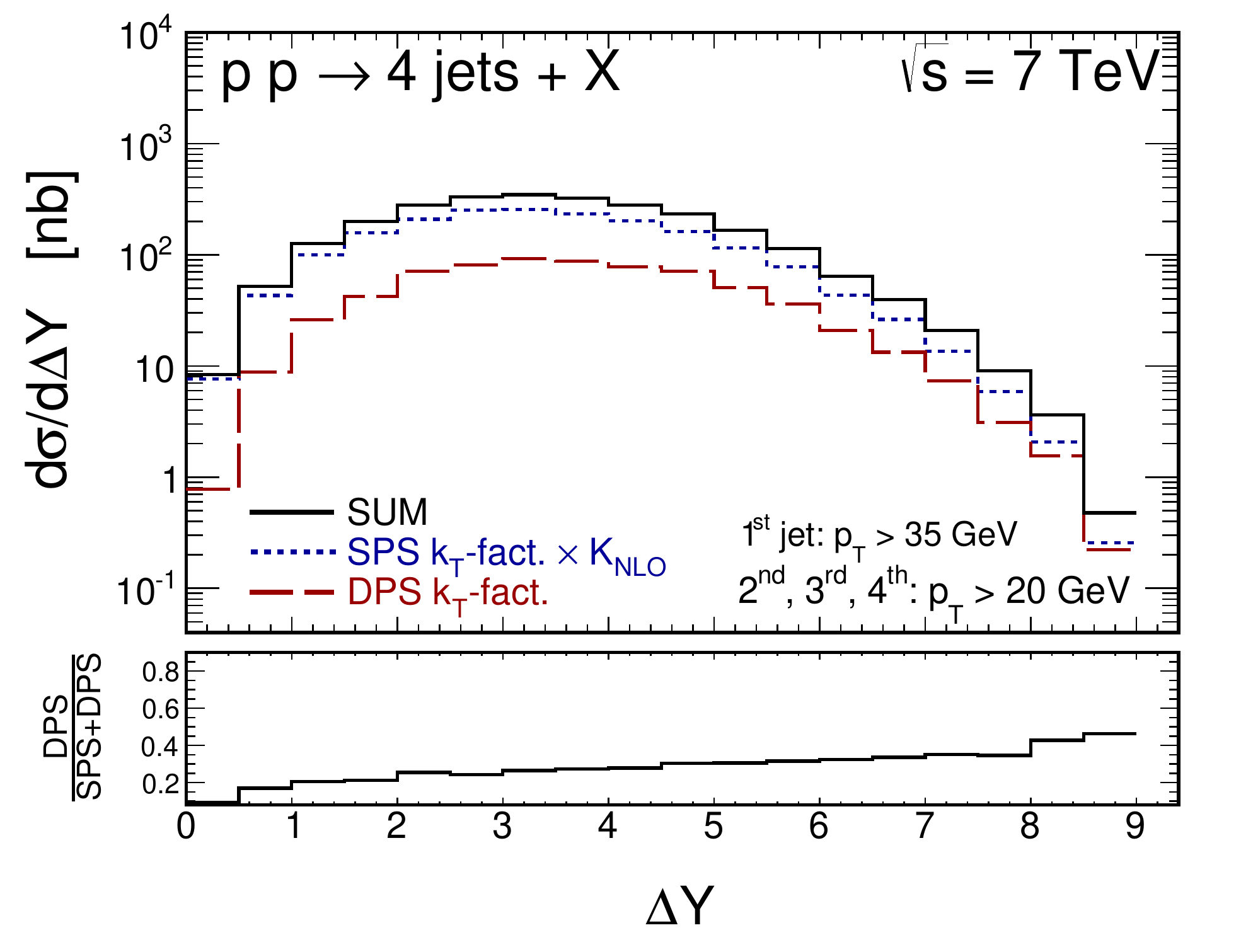}}
\end{minipage}
\hspace{0.5cm}
\begin{minipage}{0.47\textwidth}
 \centerline{\includegraphics[width=1.0\textwidth]{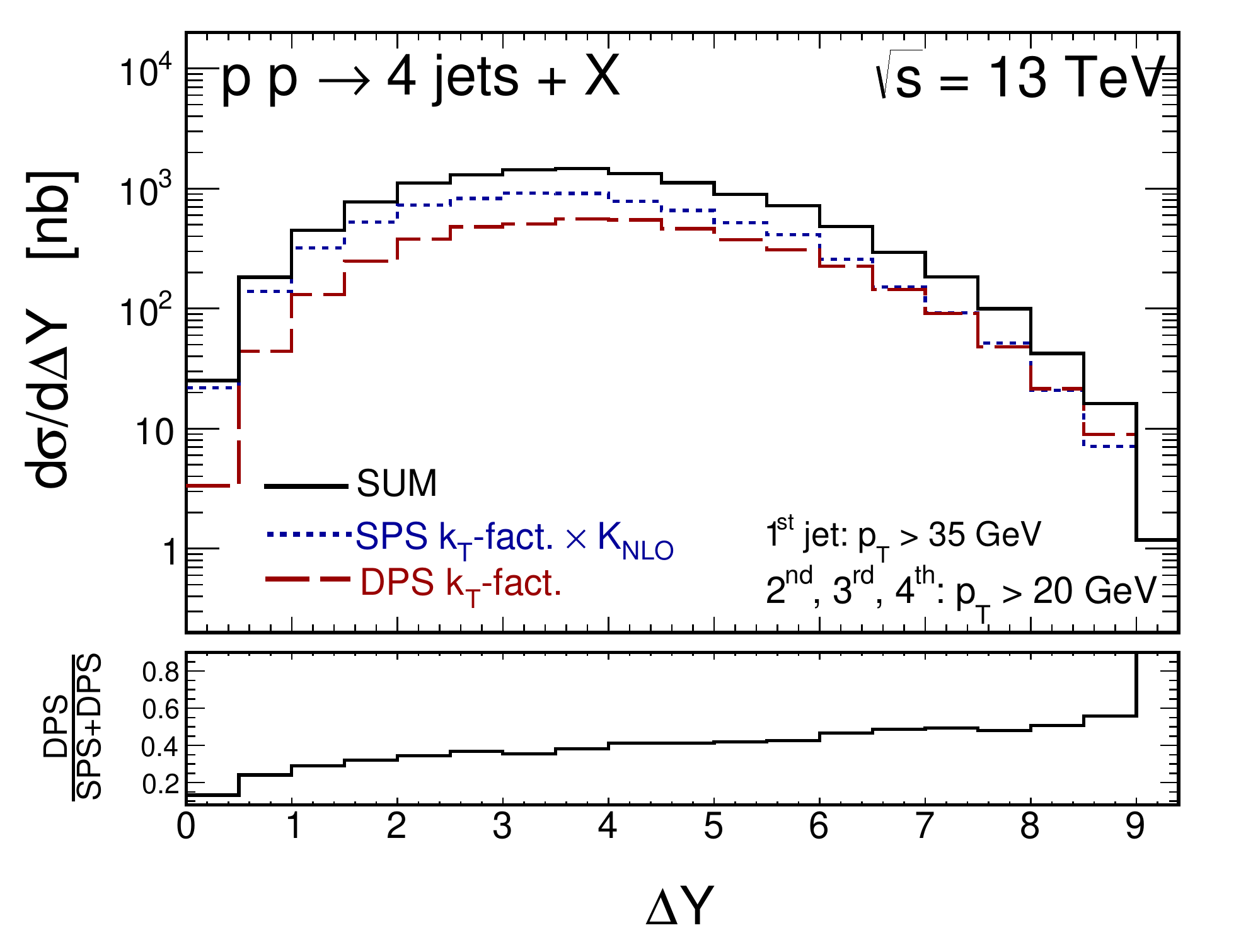}}
\end{minipage}
   \caption{
\small Distribution in rapidity distance between the most remote jets for 
the asymmetric cut  
for $\sqrt{s}$ = 7 TeV (left) and $\sqrt{s}$ = 13 TeV (right).
The SPS contribution is shown by the dotted line while 
the DPS contribution by the dashed line.
The relative contribution of DPS is shown in the extra lower panels.
 }
 \label{fig:dsig_dydiff_asymmetric}
\end{figure}

\begin{figure}[!h]
\begin{minipage}{0.47\textwidth}
 \centerline{\includegraphics[width=1.0\textwidth]{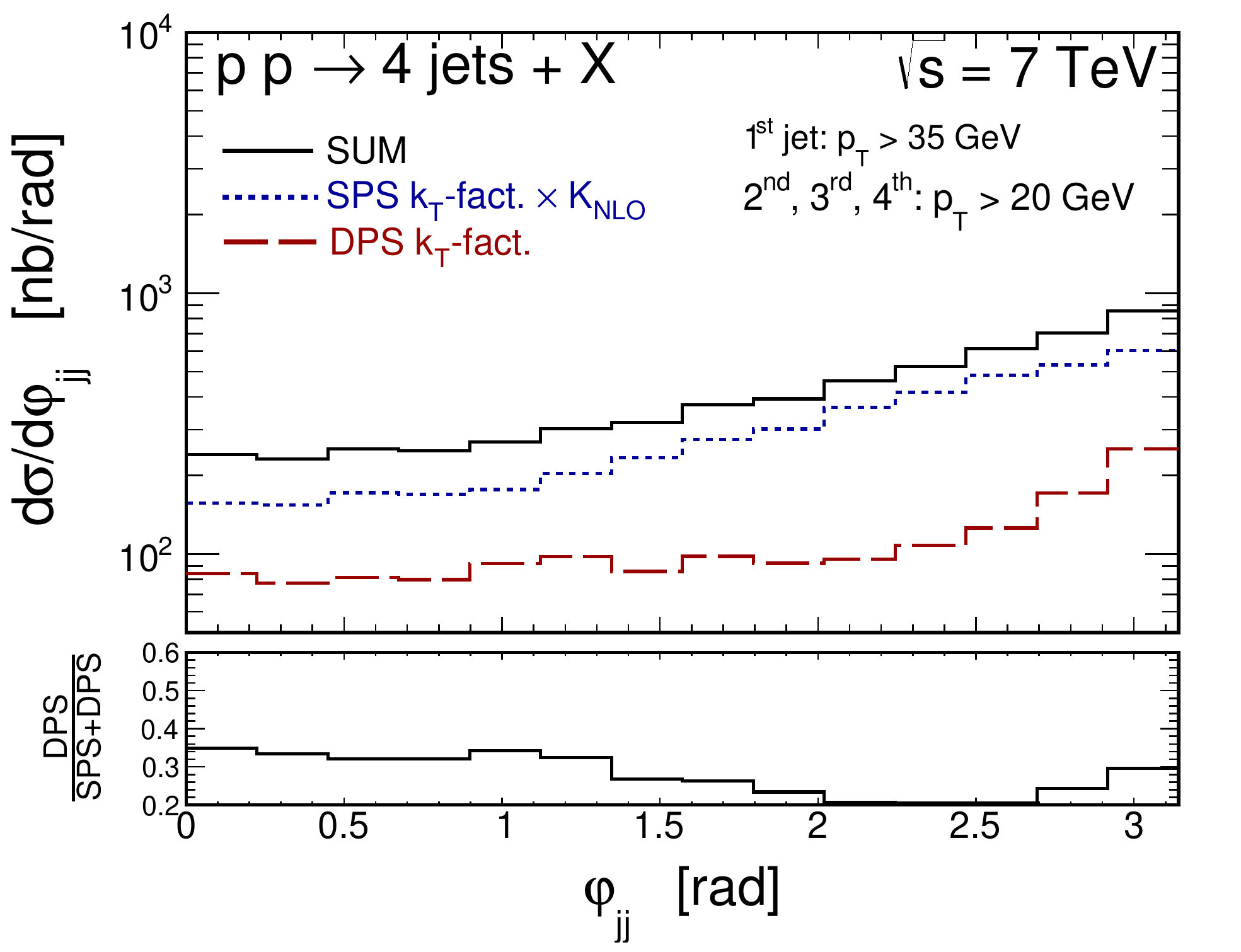}}
\end{minipage}
\hspace{0.5cm}
\begin{minipage}{0.47\textwidth}
 \centerline{\includegraphics[width=1.0\textwidth]{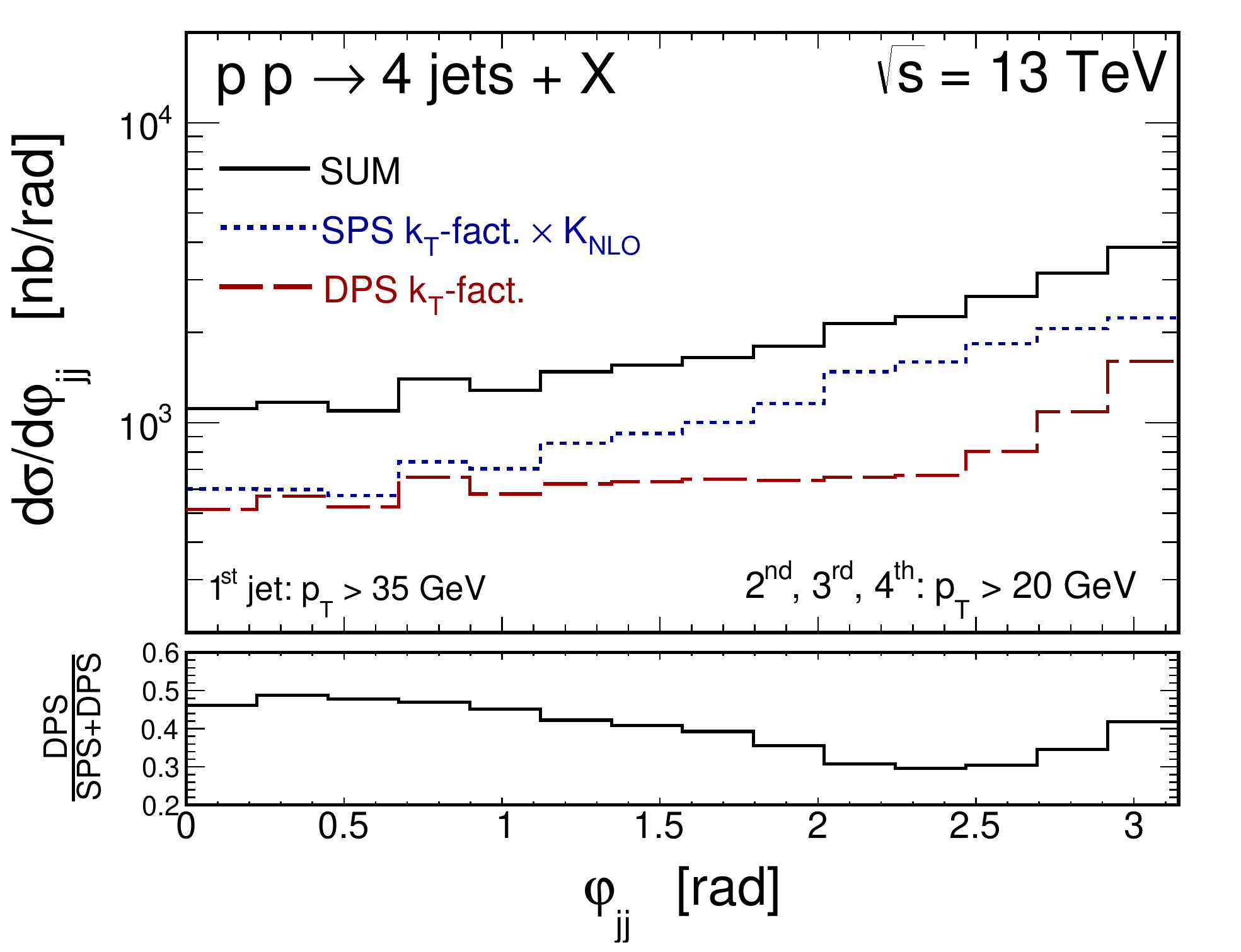}}
\end{minipage}
   \caption{
\small Distribution in relative azimuthal angle between the most remote jets for 
the asymmetric cut 
for $\sqrt{s}$ = 7 TeV (left) and $\sqrt{s}$ = 13 TeV (right).
The SPS contribution is shown by the dotted line while 
the DPS contribution by the dashed line.
The relative contribution of DPS is shown in the extra lower panels.
 }
 \label{fig:dsig_dphijj_asymmetric}
\end{figure}

\begin{figure}[t]
\begin{minipage}{0.47\textwidth}
 \centerline{\includegraphics[width=1.0\textwidth]{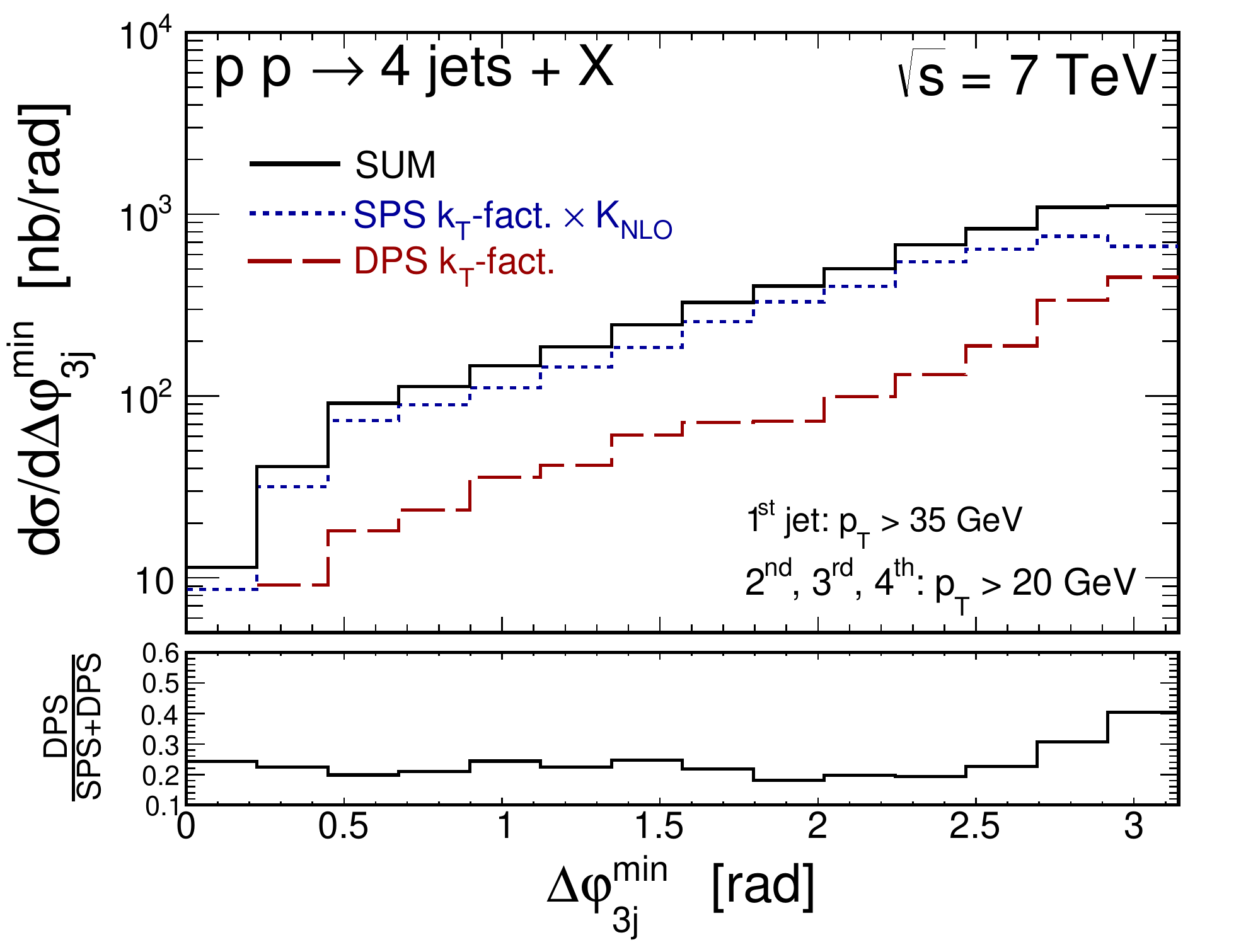}}
\end{minipage}
\hspace{0.5cm}
\begin{minipage}{0.47\textwidth}
 \centerline{\includegraphics[width=1.0\textwidth]{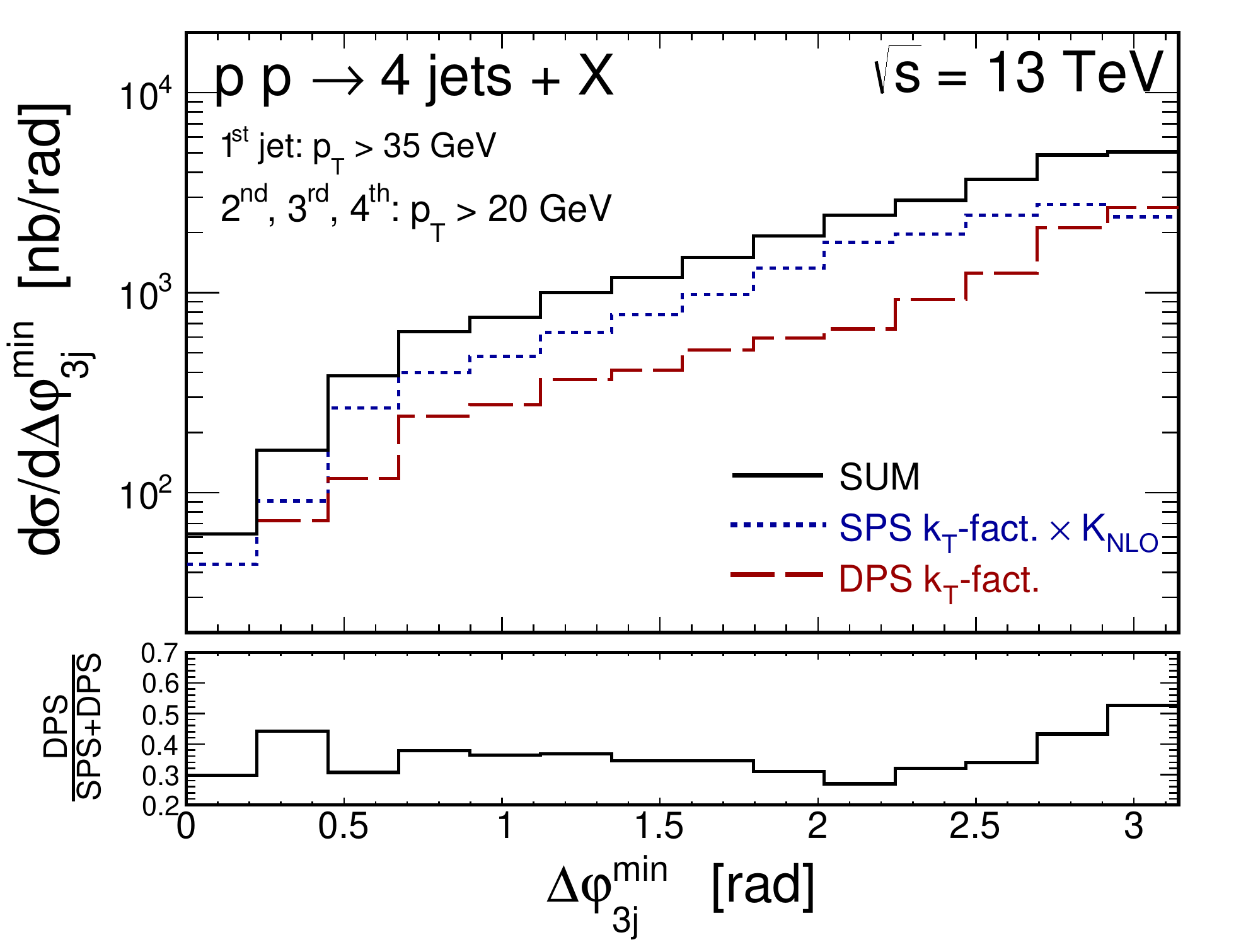}}
\end{minipage}
   \caption{
\small Distribution in $\Delta \varphi_{3j}^{min}$ angle for 
the asymmetric cut 
for $\sqrt{s}$ = 7 TeV (left) and $\sqrt{s}$ = 13 TeV (right).
The SPS contribution is shown by the dotted line while 
the DPS contribution by the dashed line.
The relative contribution of DPS is shown in the extra lower panels.
 }
 \label{fig:dsig_dphi3_asymmetric}
\end{figure}

In order for our analysis to be complete, we present here
the same variables discussed in the previous section in such an asymmetric setup.
In Figs.~\ref{fig:dsig_dy_asymmetric},~\ref{fig:dsig_dydiff_asymmetric},~\ref{fig:dsig_dphijj_asymmetric} 
and ~\ref{fig:dsig_dphi3_asymmetric} we show our predictions for the
following cuts:
$p_T > 35$ GeV for the leading jet, and $p_T > 20$ GeV for the remaining jets.
In our $k_T$-factorization framework and for these particular variables, 
the situation appears to be very similar to the situation of symmetric cuts.

\begin{figure}[!h]
\begin{minipage}{0.47\textwidth}
 \centerline{\includegraphics[width=1.0\textwidth]{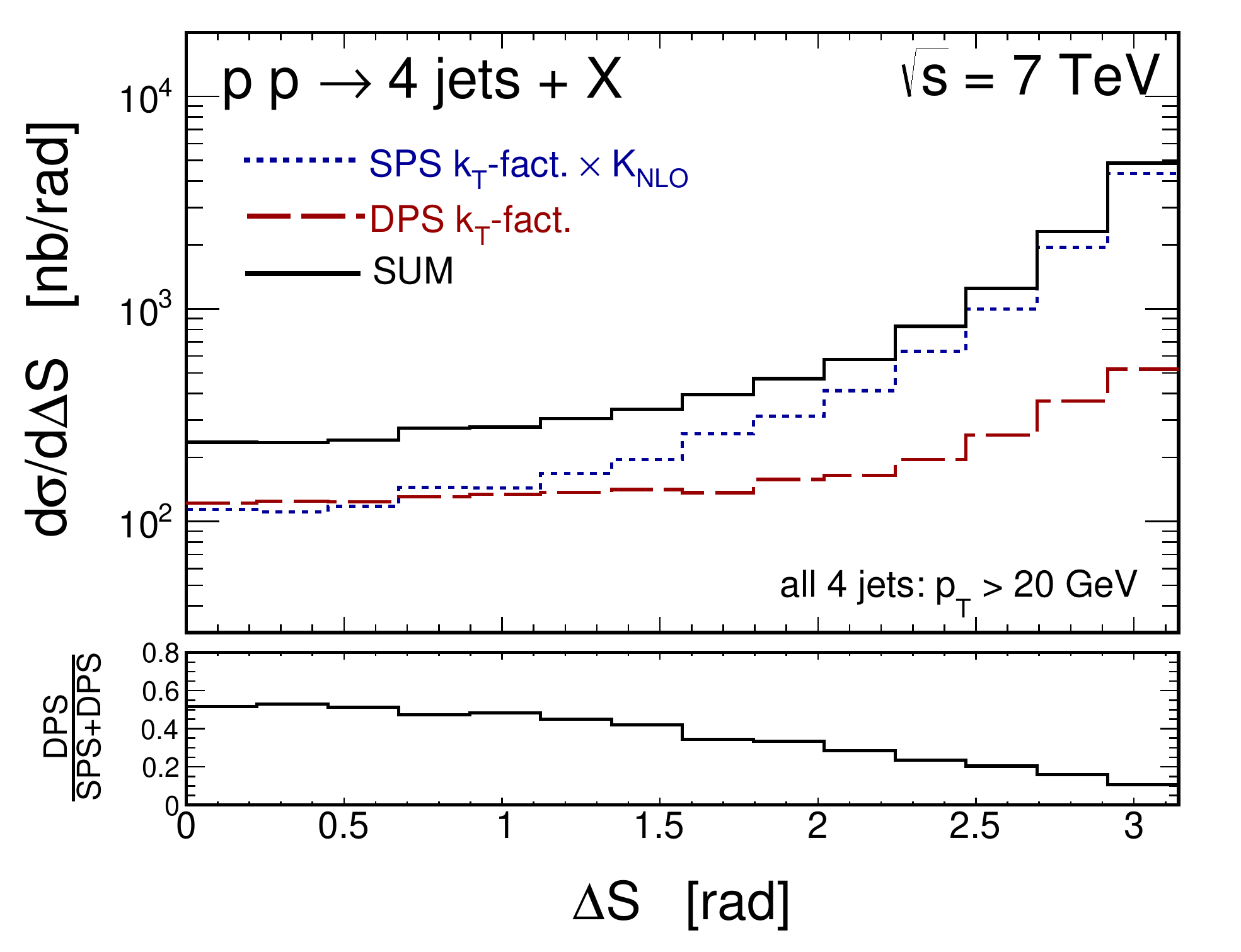}}
\end{minipage}
\hspace{0.5cm}
\begin{minipage}{0.47\textwidth}
 \centerline{\includegraphics[width=1.0\textwidth]{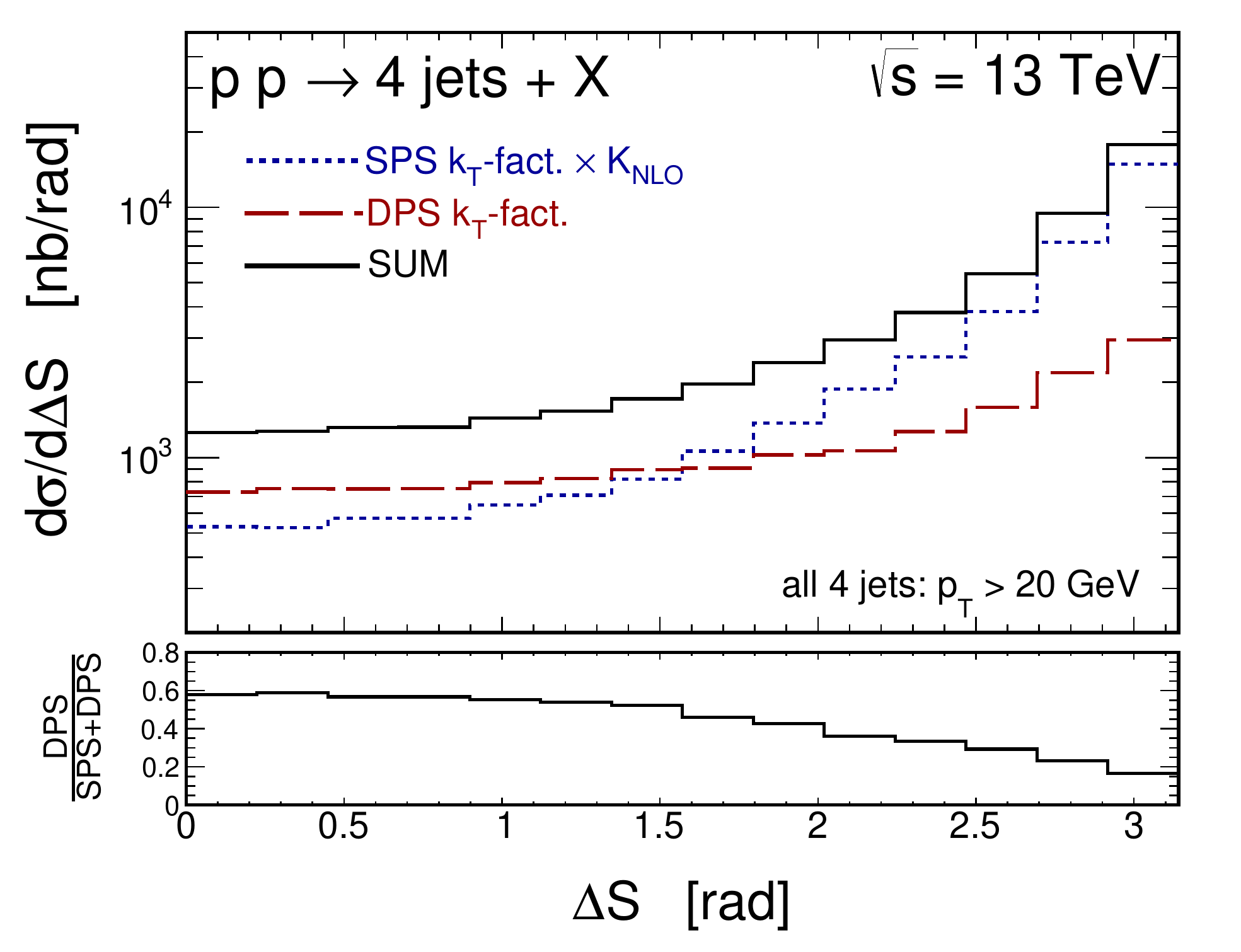}}
\end{minipage}
   \caption{
\small Distribution in $\Delta S$ for 
the asymmetric cut 
for $\sqrt{s}$ = 7 TeV (left) and $\sqrt{s}$ = 13 TeV (right).
The SPS contribution is shown by the dotted line while 
the DPS contribution by the dashed line.
The relative contribution of DPS is shown in the extra lower panels.
 }
 \label{fig:dsig_dDeltaS_asymmetric}
\end{figure}

\nopagebreak[12]
\section{Conclusions}

In the present study we have discussed how to explore DPS effects in four jet production. 
We have used results obtained in the $k_T$-factorization formalism, 
for both single-parton scattering and double-parton scattering, 
and we have discussed how to maximize their role.

Here we have shown that our approach is able to describe existing CMS data
on jet rapidity distributions and we have presented our predictions
for rapidity distributions, distribution in the distance between the most
remote jets, azimuthal angle between the most remote jets and
a new $\Delta\varphi_{3j}^{min}$ variable.

We find that, for sufficiently small cuts on the transverse momenta, 
DPS effects are enhanced relative to the SPS contribution
\begin{itemize}
\item when rapidities of jets  are large,
\item for large rapidity distances between the most remote jets,
\item for small azimuthal angles between the two jets most remote in rapidity,
\item for large values of $\Delta\varphi_{3j}^{min}$.
\end{itemize}
In general, the relative effects of DPS in the $k_T$-factorization approach are somewhat 
smaller than those found previously in the LO collinear approach. 

Both the CMS and ATLAS collaborations could perform corresponding analyses.
Future exploration of DPS effects could help in finding a new, more precise value 
for $\sigma_{eff}$ for the proton and/or finding a signal of 
a dependence of $\sigma_{eff}$ on kinematical variables. 
Such a dependence was predicted e.g.\ in a two-component model 
with perturbative-parton-splitting mechanism \cite{Gaunt:2014rua}
but has not been clearly identified experimentally yet.

\section*{Acknowledgments}
We thank Hannes Jung for feedback and many useful discussions.
The work of M.S. and K.K. have been supported by Narodowe Centrum Nauki
with Sonata Bis grant DEC-2013/10/E/ST2/00656 while R.M. and A.S.
have been supported by the Polish National Science Center grant
DEC-2014/15/B/ST2/02528. A.v.H. was supported by
a grant of National Science Center, Poland, No. 2015/17/B/ST2/01838.
M.S. also thanks the "Angelo della Riccia" foundation for support.

\bibliography{DPS_4jets_kT}
\end{document}